\documentclass[twocolumn,showpacs,prb]{revtex4}
\usepackage{graphics}
\usepackage{epsfig}

\begin{document}

\title{Resonant unidirectional and elastic scattering of surface plasmon polaritons
by high-refractive index dielectric nanoparticles}

\author{Andrey B. Evlyukhin$^{1,2}$ and Sergey I. Bozhevolnyi$^{2}$ }

\affiliation{$^1$Laser Zentrum Hannover e.V., Hollerithallee 8,
D-30419 Hannover, Germany}
 \affiliation{$^2$ Department of Technology and Innovation, University of Southern Denmark, Niels Bohrs All\'{e} 1, DK-5230 Odense, Denmark}

\date{\today}

\begin{abstract}
We consider scattering of surface plasmon polaritons (SPPs) and
light by individual high-refractive-index dielectric nanoparticles
(NPs) located on a metal (gold) substrate and supporting electric
and magnetic dipole resonances in the visible spectral range.
Numerical calculations are carried out by making use of the discrete
dipole approximation including the multipole decomposition
procedure. Extinction and scattering cross section spectra of
spheroid silicon NPs in visible and near-infrared are presented and
discussed.  The roles of the
 in-plane and out-of-plane components of  electric and magnetic
dipoles in the scattering processes are clarified and demonstrated.
It is revealed that, owing to the NP interaction with
electromagnetic fields reflected from  the substrate (that leads to
bianisotropy), the in-plane electric and magnetic dipoles can
resonantly be excited at the same wavelength. Due to this effect,
the resonant unidirectional (forward) and elastic (in-plane)
scattering of SPPs by oblate spheroid NPs can be realized within a
narrow spectral range. In the case of normal light incidence, the
bianisotropy effect can provide significant suppression of the SPP
excitation because of the destructive interference between the SPP
waves generated by induced electric and magnetic dipole moments. The
results obtained open new possibilities for the development of
SPP-based photonic components and metasurfaces, whose operation
involves resonant excitations of dielectric NPs.
\end{abstract}
\maketitle

\section{\label{sec1} INTRODUCTION}

Surface optical electromagnetic waves can be excited at the
interface between a dielectric and a metal and are called surface
plasmon polaritons (SPPs). One of the most attractive properties of
SPPs is the possibility to concentrate and channel electromagnetic
radiation using subwavelength
structures.\cite{Maier2007,Stockman2011} Experimental studies
demonstrated that ensembles of metal nanoparticles (NPs) arranged on
metal surfaces can be used to realize efficient micro-optical
elements for two-dimensional (2D) manipulation of SPP waves, such as
mirrors, beam splitters and
interferometers\cite{BozhevolnyiPRL1997,Krenn2002,Krenn2003,Stepanov2005,Evlyukhin2007,Radko2007}.
Furthermore, periodic arrays of metal NPs have been shown to exhibit
band gap properties for SPPs.\cite{Radko2009} If such an SPP
band-gap structure has narrow channels free from NPs, then SPP waves
can be confined to and guided along these channels.\cite{Radko2009}

From the theoretical point of view, the SPP scattering properties of
individual metal NPs and NP ensembles mentioned above can
efficiently be treated in the point-dipole
approximation,\cite{Bozhevolnyi2004,EvlyukhinPRB2005} with each NP
being considered as an electric dipolar scatterer and the NP dipole
moments being self-consistently determined by solving the
couple-dipole equations.\cite{Sond2003,Evlyukhin2007}
 It has been shown that electric
 dipolar NPs scatter SPP waves mainly isotropically, with only small modulation in the forward and backward
 directions with respect to the incident SPP propagation.\cite{EvlyukhinPRB2005}
 With increasing of NP sizes, higher multipole modes of NPs can
 be excited influencing and changing the SPP scattering
 diagrams\cite{Kuznetsov2009}, a feature that could be used for
 functional optimization of SPP-based components.
 It should, however, be borne in mind that metal NPs  efficiently
 absorb the electromagnetic radiation at optical
 frequencies due to large Ohmic losses.   This feature together with
 the (inelastic or out-of-plane) scattering SPP into light,
 propagating away from
 metal surface,\cite{MaradudinPRL} can significantly restrict the application
 potential of metal NP structures in 2D SPP-optics.

It is possible overcome this principal drawback of metal NPs and
nanostructures by employing an alternative approach  based on
high-refractive-index dielectric or semiconductor NPs. Due to Mie
resonances, the dielectric NPs can strongly interact with both
electric and magnetic field components of optical
waves.\cite{Bohren2008} Recently it has been shown both
theoretically\cite{EvlyukhinPRB2010, Garcia-Etxarri2011} and
experimentally\cite{EvylukhinNL2012, Kuznetsov2012} that crystalline
silicon NPs exhibit pronounced optical Mie resonances associated
with the excitation of magnetic and electric dipole (MD and ED)
modes in these NPs. Depending on NP sizes, the resonances can be
realized  within the visible or near-infrared spectral ranges, where
the silicon dielectric permittivity has a very small imaginary part
and the silicon NPs can be considered as high-refractive-index
dielectric NPs. Because silicon NPs support resonant excitation of
MD and ED modes, there appears a possibility to realize a strong
unidirectional (forward) light scattering by satisfying the
so-called Kerker
condition.\cite{Kereker,EvlyukhinPRB2010,Kuznetsov2013} Importantly,
the spectral separation between the ED and MD resonances can be
controlled by NP shapes and irradiation
conditions\cite{EvlyukhinPRB2011} that gives a possibility to
realize the resonant Kerker effect and to create novel
metamaterials.\cite{Staude2013,Yu2015,Decker2015} Resonant  optical
properties of high-refractive dielectric NPs have been suggested for
realization of new types of optical nanoantennas
\cite{Krasnok2012,Rolly2012, Schmidt2012} and ultra-high efficiency
photovoltaic cells.\cite{Spinelli2012, Brongersma2014} However,
resonant interaction of high-refractive index dielectric NPs with
SPPs have not hitherto been considered.

In this paper we concentrate on this problem and theoretically
investigate SPP and light scattering by individual silicon NPs
located on a metal (gold) substrate and supporting ED and MD
resonances in the visible spectral range. It is predicted and
numerically demonstrated that, due to NP interaction with the
electromagnetic fields reflected from  the substrate (bianisotropy
effect) the resonant unidirectional (forward) and elastic scattering
of SPPs by spheroid silicon NPs can be realized  within a narrow
spectral range and exploited for developing new SPP-based photonic
components and metasurfaces.

\begin{figure}
\begin{center}
(a)\includegraphics[scale=.100,width=14pc,height=8pc]{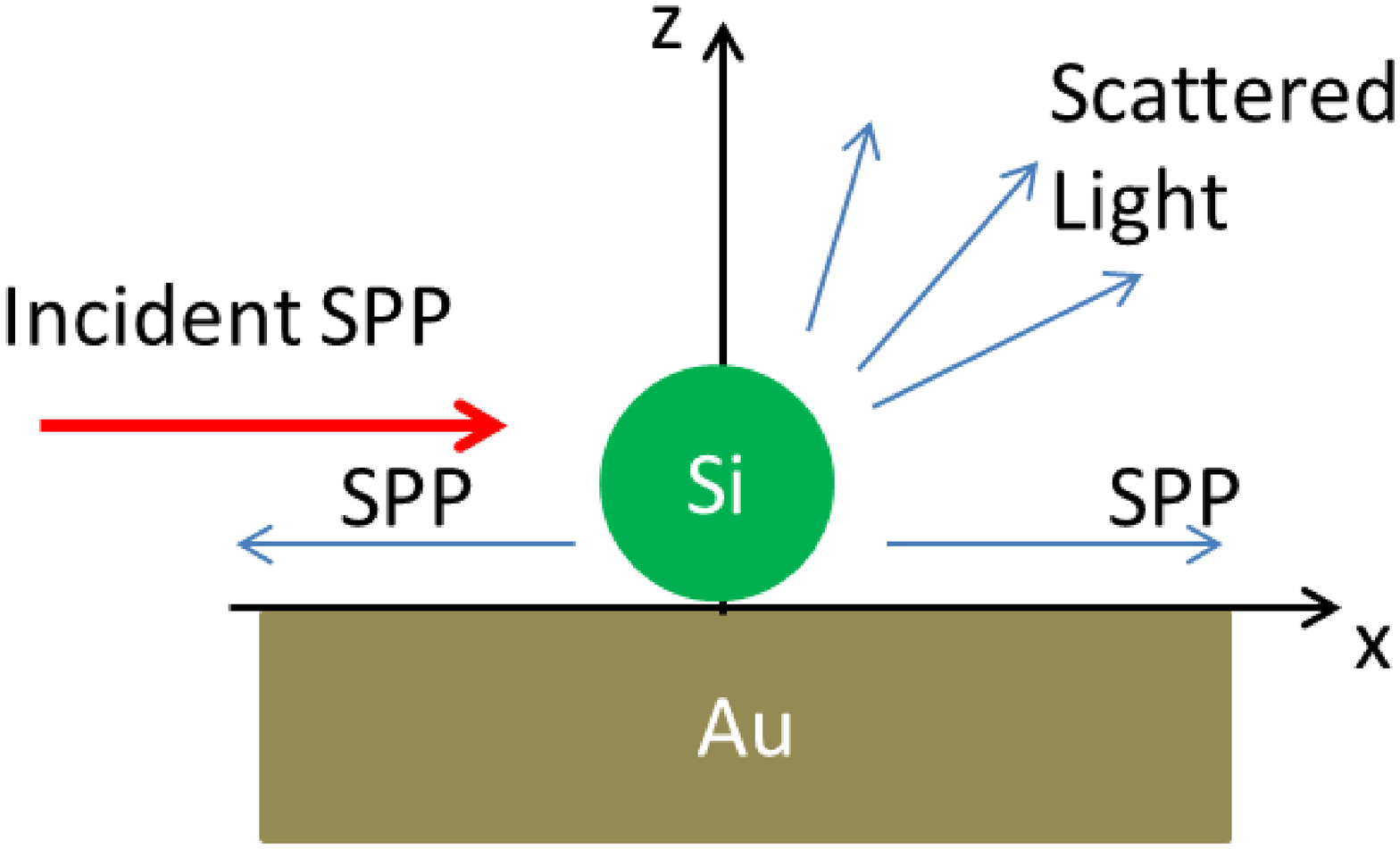}%

(b)\includegraphics[scale=.100,width=14pc,height=8pc]{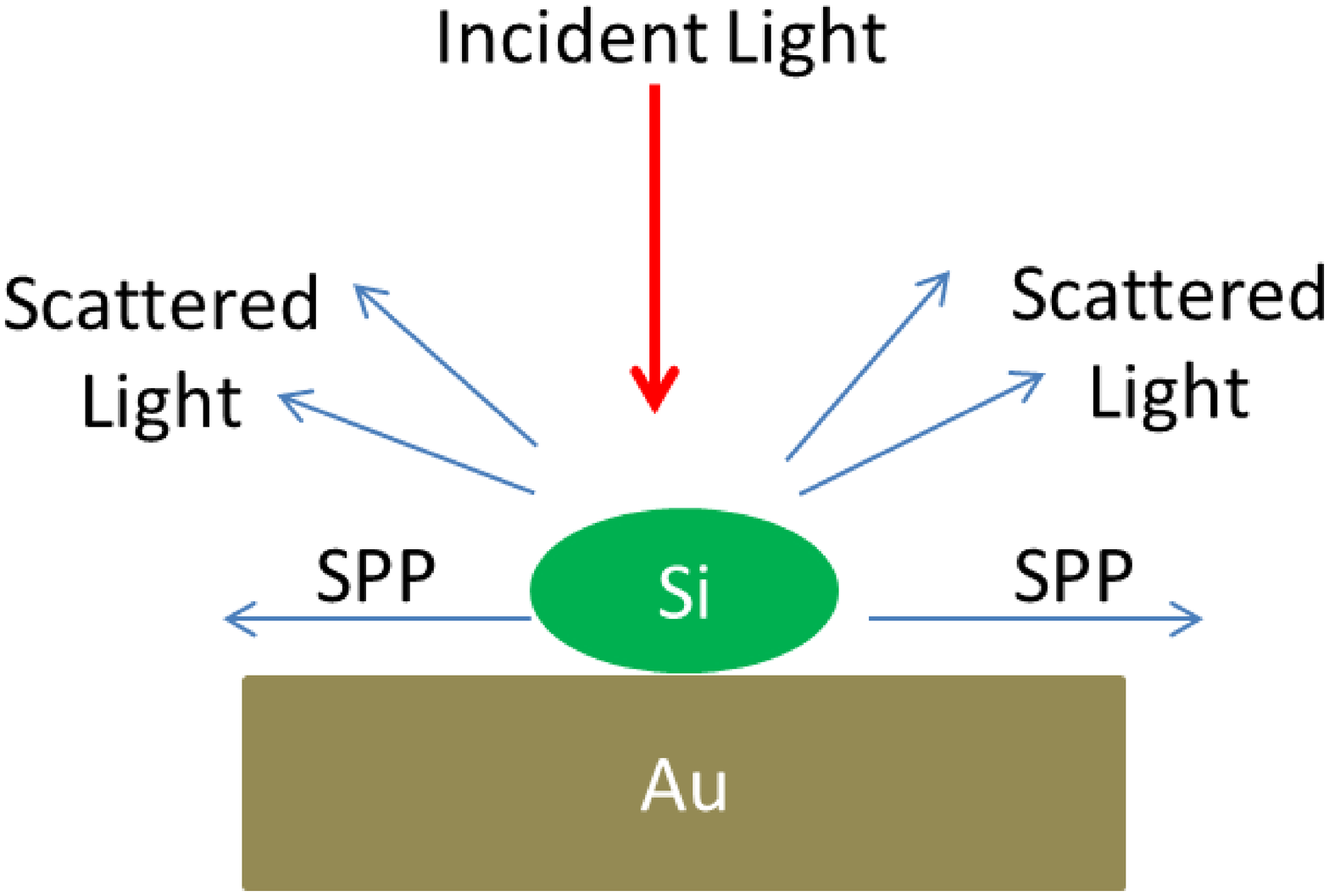}%
\end{center}
\caption{\label{F1} Schematic presentation of (a) SPP  and (b) light
scattering processes by an individual silicon NP located on a gold
substrate. Scattering channels include the SPP excitation and
scattering into light propagating away from the substrate.}
\end{figure}

\section{Theoretical background}
We assume that a silicon NP is located near a metal surface
supporting the SPP propagation (Fig.\ref{F1}). This implies that
dielectric permittivity $\varepsilon_s$ of metal satisfies the
conditions $-Re(\varepsilon_s)>\varepsilon_d>0$ and
$-Re(\varepsilon_s)>>Im(\varepsilon_s)$ in  the considered spectral
range, where $\varepsilon_d$ is the dielectric permittivity of a medium above the metal surface (in this paper $\varepsilon_d=1$).
The last inequality guaranties that the SPP propagation length is
larger than the SPP wavelength. Under these conditions, for the sake of
simplicity, we can neglect the imaginary part of $\varepsilon_s$ for
calculations of extinction and scattering cross sections
\cite{EvlyukhinPRB2007,EvlyukhinJOSAB2013}.

When an NP is illuminated either  by  an  SPP wave propagating along
the substrate surface or a light wave incident on the substrate,
there are two scattering channels: i) into SPPs and ii)
into light  propagating away from the metal surface (Fig.\ref{F1}).
The scattered electric field above the surface ($z>0$, Fig.\ref{F1})
is
\begin{equation}\label{SPP1}
{\bf E}_a({\bf r})={\bf E}_d({\bf r})+{\bf E}_r({\bf r})+{\bf
E}_{\rm SPP}({\bf r})\:,
\end{equation}
where ${\bf E}_d({\bf r})$ and ${\bf E}_r({\bf r})$ are the electric
fields of direct light waves and light reflected from metal surface,
respectively, and ${\bf E}_{\rm SPP}({\bf r})$ is the field of the
scattered SPPs. Applying the procedure of multipole decomposition in
the far-field approximation for monochromatic electromagnetic waves
(with the time dependence defined by $\exp(-i\omega t)$, where
$\omega$ is the angular frequency), the electric field ${\bf
E}_{scat}$ of scattered waves can be expresseded in the following
general form\cite{EvlyukhinJOSAB2013}
\begin{eqnarray}\label{SPP3}
{\bf E}_{scat}({\bf r})&\approx &\frac{k_0^2e^{-ik{\bf n}\cdot{\bf
r}_0}}{\varepsilon_0} \hat S({\bf r})\left\{{\bf p}-
\frac{k}{\omega}[{\bf{n}}\times{\bf m}]\right.\nonumber\\
&-&\frac{ik}{6}\:\hat Q{\bf{n}}-\frac{k^2}{6}\:\hat
O({\bf{n}}{\bf{n}})+\left.\frac{ik^2}{2\omega}[{\bf{n}}\times\hat
M{\bf{n}}]\right\}\:.\nonumber\\
\end{eqnarray}
where  $k_0$ is the wave number of light, $\varepsilon_0$ is the
vacuum dielectric permittivity, ${\bf p}$ and ${\bf m}$ are the
electric and MD moments of the scatterer, respectively, $\hat Q$ and
$\hat M$ are the electric and magnetic quadrupole tensors of the
scatterer, respectively, and $\hat O$ is the electric octupole
tensor. The multipole moments are calculated (and consequently
located) at the point ${\bf r}_0$. The value $k$, vector ${\bf{n}}$
and the tensor $\hat{S}({\bf r})$ are determined by  the scattered
waves (\ref{SPP1}).  For ${\bf E}_d({\bf r})$:
$k=k_d=k_0\sqrt{\varepsilon_d}$; ${\bf{n}}={\bf r}/|{\bf r}|$. For
${\bf E}_r({\bf r})$: $k=k_d=k_0\sqrt{\varepsilon_d}$;
${\bf{n}}={\bf r'}/|{\bf r'}|$, where ${\bf r'}=(x,y,-z)$.  For
${\bf E}_{\rm SPP}({\bf r})$:
$k=k_S=k_0\sqrt{\varepsilon_s\varepsilon_d/(\varepsilon_s+\varepsilon_d)}$;
 ${\bf n}=(x/\rho, y/\rho,
-ia)$, $\rho=\sqrt{x^2+y^2}$,
$a=\sqrt{-\varepsilon_d/\varepsilon_s}$. The multipole moments of
the NP are calculated using the induced polarization ${\bf P}({\bf
r})=\varepsilon_0(\varepsilon_p-\varepsilon_d){\bf E}({\bf r})$
inside the NP, where ${\bf E}({\bf r})$ is the total electric field
inside the NP and $\varepsilon_p$ is its dielectric permittivity.
\begin{equation}\label{ed}
{\bf p}=\int_{V_S}{\bf P}({\bf r}')d{\bf r}'
\end{equation}
is the ED moment,
\begin{eqnarray}\label{Q2}
\hat Q\equiv\hat Q({\bf r}_0)&=&3\int_{V_S}(\Delta{\bf r}{\bf
P}({\bf r}')+{\bf P}({\bf r}')\Delta{\bf r})d{\bf r}',
\end{eqnarray}
\begin{equation}\label{md}
{\bf m}\equiv{\bf m}({\bf
r}_0)=-\frac{i\omega}{2}\int_{V_S}[\Delta{\bf r}\times{\bf P}({\bf
r}')]d{\bf r}',
\end{equation}
are the electric quadrupole tensor, the MD moment, where $V_S$ is
the NP's volume, $\Delta{\bf r}={\bf r}'-{\bf r}_0$ (it is
convenient to choose the point ${\bf r}_0$ at the NPs's center of
mass \cite{EvlyukhinPRB2011}). Note, that $\bf nn$ and $\Delta{\bf
r}{\bf P}$  represent the dyadic (outer) products between the corresponding
vectors. Expressions for higher-order multipole moments and
electric fields generated (scattered) by the corresponding multipole
moments  can be found elsewhere.\cite{EvlyukhinJOSAB2013}

\subsection{Scattering and extinction cross-sections}

Let us first consider the scattering cross sections. Since
scattered light waves propagate away from the metal surface, their
interference with scattered SPP waves propagating along the surface
is negligible weak in the far-filed zone. As a result the total
scattering cross section $\sigma_{scat}$ can be represented as the
sum of two parts $\sigma_{\rm SPP}$ and $\sigma_{\rm Light}$
corresponding to the scattering into SPP and into light propagating
away from the surface,
respectively:\cite{EvlyukhinPRB2005,EvlyukhinPRB2007}
\begin{equation}
\sigma_{\rm SPP}=\frac{P_{\rm SPP}}{P_{\rm in}}\:,  \quad {\rm and}
\quad \sigma_{\rm Light}=\frac{P_{\rm L}}{P_{\rm in}}\:,
\end{equation}
where ${P_{\rm SPP}}$ and $P_{\rm L}$ are the scattered powers into
SPP and light, respectively, $P_{\rm in}$ is the incident wave power. For the incidence of a light plane wave: $P_{\rm
in}=\sqrt{\varepsilon_0\varepsilon_d/\mu_0}|{\bf E}_0|^2/2$, and in the case of incident SPP plane wave:
$$
P_{\rm
in}=\frac{1}{2k_0}\sqrt{\frac{\varepsilon_0}{\mu_0}}\frac{1-a^2}{2a}\:
(1-a^4)|{\bf E}_0^{\rm SPP}|^2\:,
$$
where ${\bf E}_0$ and ${\bf E}_0^{\rm SPP}$ are the electric
amplitudes of the incident light plane wave and the normal
(out-of-plane) component of incident SPP plane wave, respectively.
Note that, in the case of SPP incidence,   $P_{\rm in}$ is determined
per unit length.\cite{MaradudinPRL,EvlyukhinPRB2005}

The power of scattered SPP propagating into a in-plane angle
$[\varphi, \varphi+d\varphi]$ (the polar coordinate system) is
given by the expression:\cite{EvlyukhinJOSAB2013}
\begin{equation}\label{power_p_m}
P(\varphi)d\varphi=\sqrt{\frac{\varepsilon_0}{\mu_0}}\frac{(1-a^2)(1-a^4)}{4ak_0}|E_{{\rm
SPP}z}|^2\rho d\varphi\:,
\end{equation}
where the z-component (out-of-plane) of the scattered SPP electric
field $E_{{\rm SPP}z}$ is taken  in the domain above metal surface
just on metal surface ($z=0$). Note that the expression
(\ref{power_p_m}) was obtained by neglecting the imaginary part of
metal permittivity $\varepsilon_s$. The total SPP power can be written down as follows:
\begin{equation}
P_{\rm
SPP}=\int_{0}^{2\pi}P(\varphi)d\varphi=\int_{-\pi/2}^{\pi/2}P(\varphi)d\varphi+\int_{\pi/2}^{3\pi/2}P(\varphi)d\varphi\:,
\end{equation}
where the first and second integrals from the right hand side are
the "Forward" $F_{\rm SPP}$ and "Backward" $B_{\rm SPP}$ scattering
powers, respectively (here we
 suppose that the incident SPP plane wave propagates along the positive
direction of $x$-axis as shown in Fig \ref{F1}a).

 Using the electric field $E_{{\rm SPP}z}$
as a superposition of multipole fields (\ref{SPP3}),
the contribution of different multipoles in the scattered SPP power can be
evaluated. In the electric and MD approximation one can
write\cite{EvlyukhinJOSAB2013}
\begin{equation}\label{ESPPZ}
E_{{\rm SPP}z}\sim
p_z+iap_\rho-(1-a^2)\sqrt{\mu_0\varepsilon_0}\:\frac{k_S}{k_0}\:m_\varphi\:,
\end{equation}
where $p_z$ and $p_\rho$ are the out-of-plane and in-plane
components of ED moment $\bf p$, respectively,  and $m_\varphi$ is
the in-plane component of the MD moment $\bf m$. The in-plane
components are connected with the Cartesian component of the
corresponding dipole moments: $p_\rho=p_x\cos\varphi+p_y\sin\varphi$
and $m_\varphi=-m_x\sin\varphi+m_y\cos\varphi$. It is important to
note that $i)$ the contribution of the in-plane component of ED
moment $p_\rho$ is proportional to the parameter $a$, which is very small ($a<<1$) in the spectral region where SPPs can be excited
and $ii)$ the out-of-plane component $m_z$ of MD does not
 excite SPP on metal surfaces.

The scattered power into the light waves propagating away from the
metal surface can be calculated from\cite{EvlyukhinJOSAB2013}
\begin{equation}\label{PL}
P_{\rm
L}=\frac{1}{2}\sqrt{\frac{\varepsilon_0\varepsilon_d}{\mu_0}}\int_0^{\pi/2}\int_0^{2\pi}|{\bf
E}_d+{\bf E}_r|^2r^2\sin\theta{\rm d}\theta{\rm d}\varphi\:,
\end{equation}
where $\varphi$ and $\theta$ are the azimutal and polar angles of
the spherical coordinate system, respectively. Note that the
expression
\begin{equation}
\sigma(\varphi,\theta)=\frac{1}{2P_{in}}\sqrt{\frac{\varepsilon_0\varepsilon_d}{\mu_0}}|{\bf
E}_d+{\bf E}_r|^2r^2
\end{equation}
is the differential cross section of the scattering into the light
propagating away from the substrate surface. This value describes
the angular distribution of the scattered light. In the electric and
MD approximation one can write\cite{EvlyukhinJOSAB2013}
\begin{eqnarray}\label{EL1}
({\bf E}_d+{\bf E}_r)_\varphi&\sim&[r^{(s)}e^{ik_d2z_0\cos\theta}+1]\nonumber\\
&&\times(p_\varphi+\sqrt{\mu_0\varepsilon_0\varepsilon_d}\:m_z\sin\theta)\\
&&+[r^{(s)}e^{ik_d2z_0\cos\theta}-1]\sqrt{\mu_0\varepsilon_0\varepsilon_d}\:m_\rho\cos\theta\:,\nonumber
\end{eqnarray}
and
\begin{eqnarray}\label{EL2}
({\bf E}_d+{\bf E}_r)_\theta&\sim&[r^{(p)}e^{ik_d2z_0\cos\theta}+1]\nonumber\\
&&\times(\sqrt{\mu_0\varepsilon_0\varepsilon_d}\:m_\varphi-p_z\sin\theta)\\
&&+[1-r^{(p)}e^{ik_d2z_0\cos\theta}]\:p_\rho\cos\theta\:,\nonumber
\end{eqnarray}
where $r^{(s)}$ and $r^{(p)}$ are the reflection coefficients for
$s$- and $p$-polarized waves, respectively,\cite{EvlyukhinJOSAB2013}
$z_0$ is the position of the dipole moments, $p_\rho$ and
$p_\varphi=-p_x\sin\varphi+p_y\cos\varphi$ are the in-plane
components of the particle ED moment,
$m_\rho=m_x\cos\varphi+m_y\sin\varphi$. In contrast to (\ref{ESPPZ})
contributions of the ED in-plane components in electric field
(\ref{EL1}) and (\ref{EL2}) are not proportional to the small
parameter $a$.

The  extinction (scattering+apbsorption) cross-section
$\sigma_{ext}$ can be evaluated by
\cite{Draine1988,Bozhevolnyi2004}
\begin{equation}\label{ext1}
\sigma_{\rm ext}=\frac{P_{\rm ext}}{P_{\rm
in}}=\frac{\omega}{2P_{\rm in}}\:{\rm Im}\int_{V_S}{\bf E}_0^*({\bf
r})\cdot{\bf P}({\bf r})d{\bf r}\:,
\end{equation}
where $P_{\rm ext}$ is the extinction power,  ${\bf E}_0({\bf r})$
is the electric field of the incident (external) electromagnetic
waves (the field that we would have if the scattering object was not
there). The multipole decomposition of the extinction cross section
is presented elsewhere.\cite{EvlyukhinJOSAB2013} In the case of SPP
plane wave incidence (Fig. \ref{F1}a)
\begin{equation}
{\bf E}_0({\bf r})={ E}^{\rm SPP}e^{ik_{S}x-ak_{S}z}(-ia\hat x, 0,
\hat z)\:, \quad z>0\:,
\end{equation}
where ${ E}^{\rm SPP}$ is the amplitude magnitude of electric field.
In the case of light plane wave incidence
\begin{equation}
{\bf E}_0({\bf r})=e^{ik_dx\sin\beta}({\bf E}_{0}^{\rm
in}e^{-ik_dz\cos\beta}+R{\bf E}_{0}^{\rm ref}e^{ik_dz\cos\beta})\:,
\end{equation}
where $\beta$ is the incident angle, ${\bf E}_{0}^{\rm in}$ and
${\bf E}_{0}^{\rm ref}$ are the amplitude vectors of the incident
and reflected waves, respectively,  $R$ is the reflection
coefficient depending on the wave polarization. Here we suppose that
the light incident plane is the $xz$-plane (Fig. \ref{F1}b). In the
case of $TE$-polarization ${\bf E}_{0}^{\rm in}={\bf E}_{0}^{\rm
ref}$ and
$$
R=\frac{\cos\beta-\sqrt{(\varepsilon_s/\varepsilon_d)-\sin^2\beta}}{\cos\beta+\sqrt{(\varepsilon_s/\varepsilon_d)-\sin^2\beta}}\:.
$$
In the case of $TM$-polarization  $E_{0x}^{\rm in}=-E_{0x}^{\rm
ref}$ and $E_{0z}^{\rm in}=E_{0z}^{\rm ref}$, and
$$
R=\frac{\varepsilon_s\cos\beta-\varepsilon_d\sqrt{(\varepsilon_s/\varepsilon_d)-\sin^2\beta}}
{\varepsilon_s\cos\beta+\varepsilon_d\sqrt{(\varepsilon_s/\varepsilon_d)-\sin^2\beta}}\:.
$$
The magnetic field  ${\bf H}_0$ can be found using the Maxweel
equation $\nabla\times{\bf E}_0=ik_0\sqrt{\mu_0/\varepsilon_0}{\bf
H}_0$.

\section{\label{Results} Results and discussions}

Our numerical calculations are carried out using the discrete dipole
approximation (DDA)\cite{Draine1988} including the procedure of the
multipole decomposition.\cite{EvlyukhinJOSAB2013} We consider
first a spherical 95-nm-radius silicon NP placed near a
gold surface (the space gap between the NP and the substrate is
equal to 5 nm). The studied system together with the chosen coordinate
system are shown in Fig.\ref{F1}a. Dielectric constant for gold and
crystalline silicon are taken from the tables in Refs. {JC1972} and {Palik}, respectively.
Importantly, for gold ${\rm Re}\:a>>{\rm Im}\:a$ and $a\approx{\rm
Re}\:a\leq 0.3$ and $k_S/k_0\approx {\rm Re}\:k_S/k_0\approx1$ in
the considered optical range. This will be used in the following when conducting estimations.

\begin{figure}
\begin{center}
(a) \includegraphics[scale=.100,width=17pc,height=13pc]{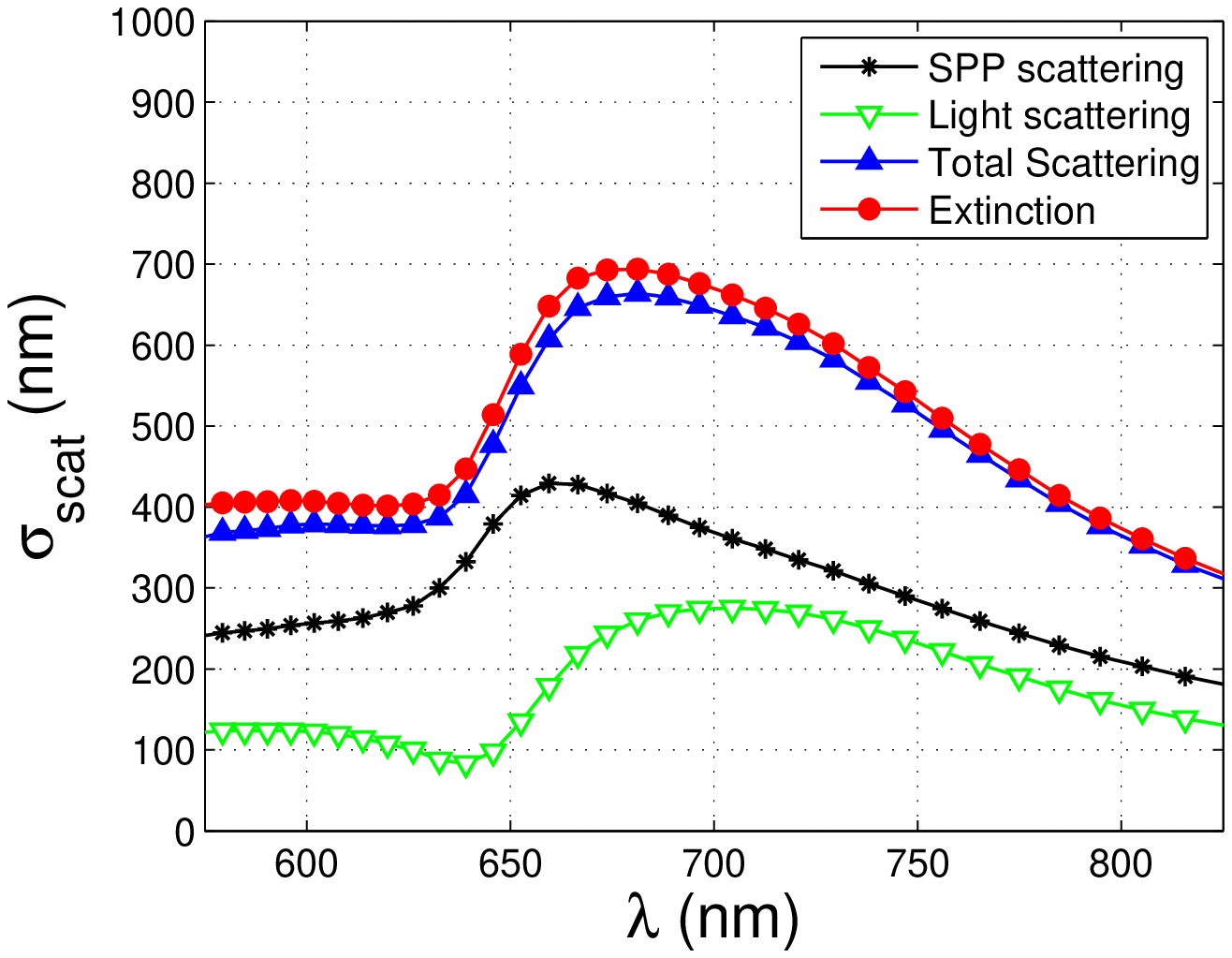}%

(b) \includegraphics[scale=.100,width=17pc,height=13pc]{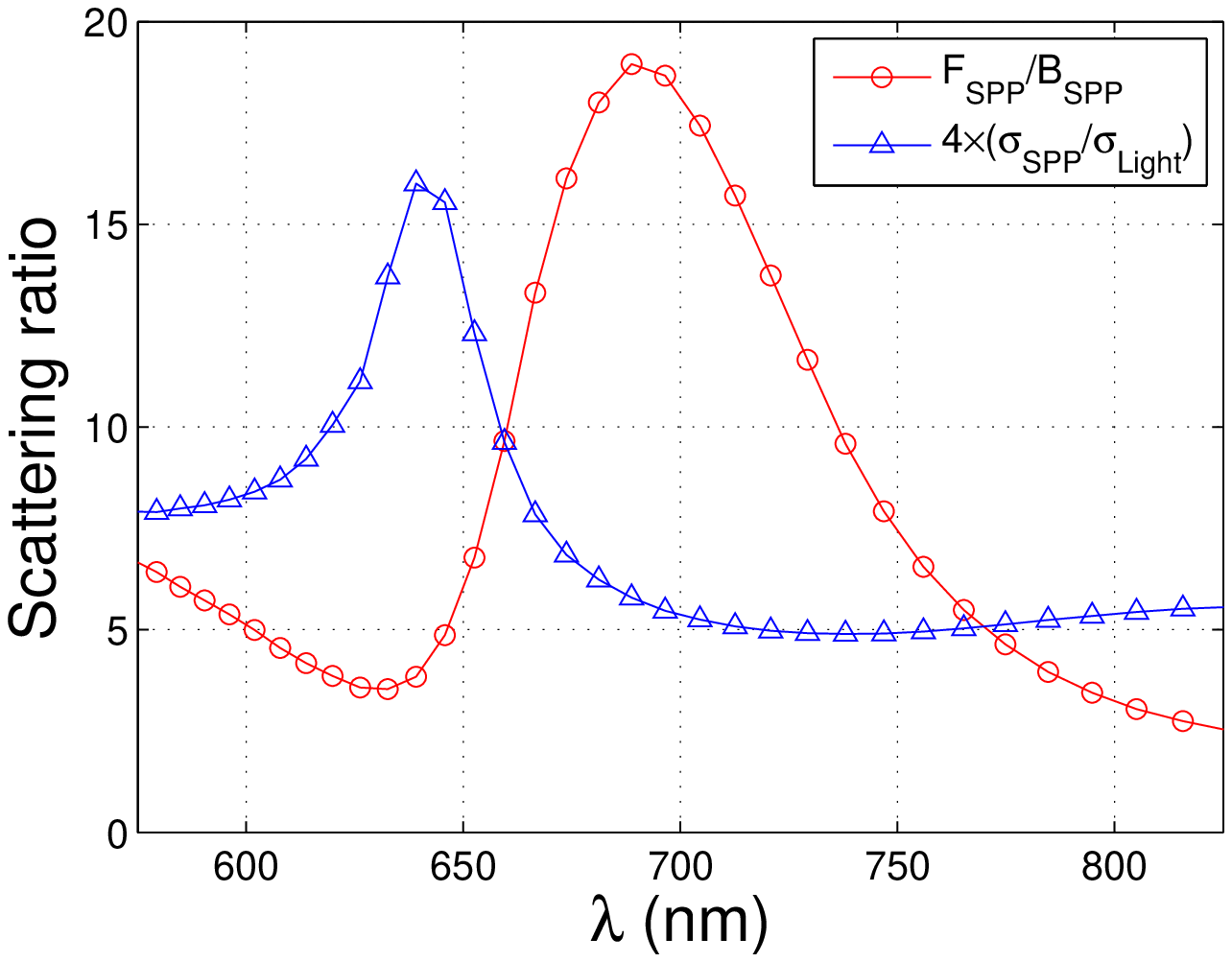}%
\end{center}
\caption{\label{F2} (a) Extinction and scattering cross-section
spectra of a spherical silicon NP (with the radius of 95 nm) located
on a gold surface and irradiated by a SPP plane wave. SPP (Light)
scattering corresponds to $\sigma_{\rm SPP}$ ($\sigma_{\rm Light}$).
(b) Corresponding spectral dependence of the ratios $\sigma_{\rm
SPP}/\sigma_{\rm Light}$ and $F_{\rm SPP}/B_{\rm SPP}$. For
convenient presentation   the ratio $\sigma_{\rm SPP}/\sigma_{\rm
Light}$ was multiplied by the factor 4. }
\end{figure}

\subsection{SPP scattering}

\begin{figure}
\begin{center}
\includegraphics[scale=.100,width=17pc,height=13pc]{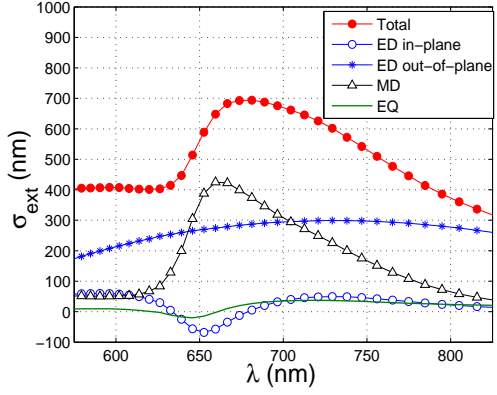}%
\end{center}
\caption{\label{F3} Extinction cross-section spectra of a spherical
silicon NP (with the radius of 95 nm) located on a gold surface.
The plot shows different multipole contributions to the total
extinction cross section: ED electric dipole; MD magnetic dipole; EQ
electric quadrupole. }
\end{figure}

\begin{figure}
\begin{center}
\includegraphics[scale=.100,width=17pc,height=13pc]{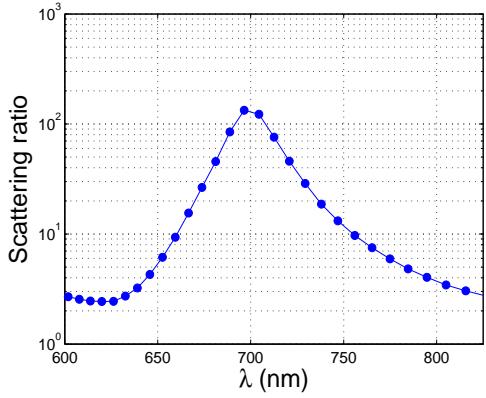}%
\end{center}
\caption{\label{F4} Spectral dependence of the ratio (\ref{rr}) for
the parameters as in Fig. \ref{F2}. }
\end{figure}

Scattering and extinction  cross-sections for the case of SPP scattering by the
spherical silicon NP are shown in Fig.\ref{F2}a. Small differences
between the scattering and extinction cross sections indicate on
very weak absorption of electromagnetic energy by the NP. The
partial scattering cross-sections for the two scattering channels
are also presented in Fig. \ref{F2}a. In the whole considered spectral
range, the (elastic) SPP scattering into SPPs is more efficient than the (inelastic) SPP scattering into light.
Especially, this concerns the spectral vicinity of the wavelength of 645 nm, at which the ratio between $\sigma_{\rm
SPP}$ and $\sigma_{\rm Light}$ is at its maximum and equals to 4 (blue
curve in Fig.\ref{F2}b). Note that this maximum of the elastic SPP
scattering is realized outside the resonance of the total scattering
cross-section and outside the spectral region, within which
the forward SPP-into-SPP scattering dominates over the backward one resulting in the unidirectional SPP scattering (red
curve in Fig.\ref{F2}b). For physical clarification  of these
scattering features, we employ a multipole analysis of SPP scattering
by NPs on a plane surface.\cite{EvlyukhinJOSAB2013} From the
multipole decomposition presented in Fig. \ref{F3} one can see that
the extinction is basically determined by contributions of the
out-of-plane ED ($p_z$) and in-plane MD ($m_y$) moments of the NP.
Contributions of the in-plane ED ($p_x$) and electric quadrupole
moments in the extinction cross-section are weak. The broad
peak is associated with resonant excitation of $m_y$ at the
wavelength of 670 nm and the nonresonant contribution from $p_z$. So
that the resonant forward (unidirectional) scattering of the
incident SPP into SPPs (Fig. \ref{F2}b) is explained by interference
effect between electromagnetic fields generated  by $m_y$ and $p_z$.
From (\ref{ESPPZ}) neglecting contribution of $ap_x$ and taking into
account that $a^2<<1$ and $k_S/k_0\approx 1$ one can write
\begin{equation}
E_{{\rm SPP}z}\sim p_z-\sqrt{\mu_0\varepsilon_0}\:m_y\cos\varphi\:.
\end{equation}
At the condition of the forward (backward) scattering (with respect
to the direction of incident SPP) $\varphi=0 (\pi)$. Therefore the
ratio between the elastic forward and backward SPP scattering can be
estimated as
\begin{equation}\label{rr}
\frac{|p_z-\sqrt{\mu_0\varepsilon_0}\:m_y|^2}{|p_z+
\sqrt{\mu_0\varepsilon_0}\:m_y|^2}\:.
\end{equation}
The spectral  maximum of the ratio (\ref{rr}) shown in
 Fig. \ref{F4} is in good agreement with the corresponding maximum for
 $\rm F_{\rm SPP}/\rm B_{\rm SPP}$ in Fig. \ref{F2}b confirming a
 very small role of the $p_x$ component in the unidirectional
 SPP scattering.

\begin{figure}
\begin{center}
\includegraphics[scale=.100,width=17pc,height=13pc]{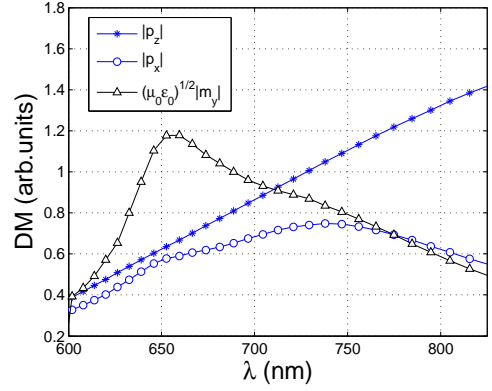}%
\end{center}
\caption{\label{F5} The magnitudes of the in-plane $p_x$ and
out-of-plane $p_z$ components of the ED moment and the in-plane
$m_y$ component of the MD moment of spherical silicon NP irradiated
by a SPP plane wave. The parameters are as in Fig. \ref{F2}.}
\end{figure}

\begin{figure}
\begin{center}
(a)\includegraphics[scale=.100,width=17pc,height=13pc]{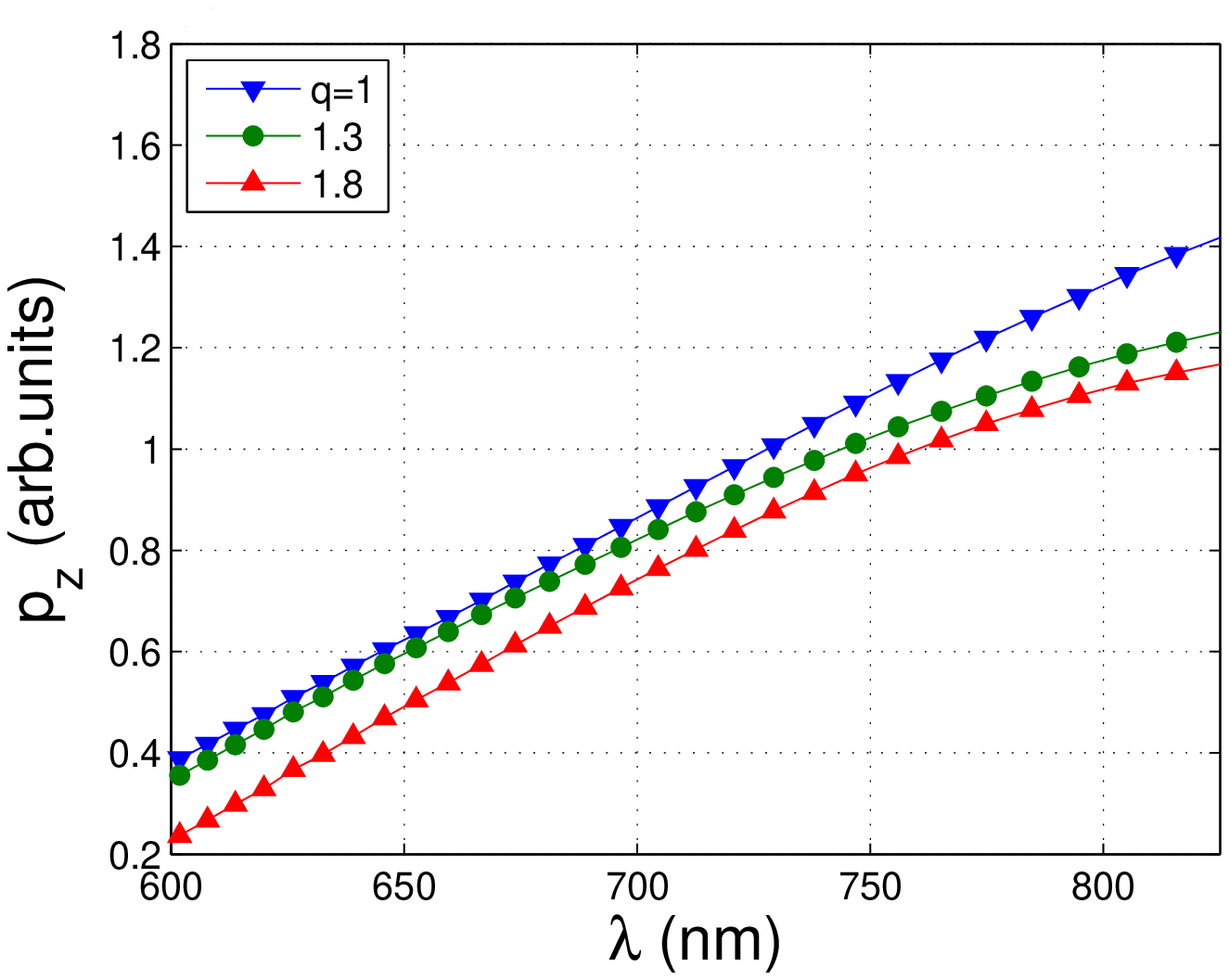}%

(b)\includegraphics[scale=.100,width=17pc,height=13pc]{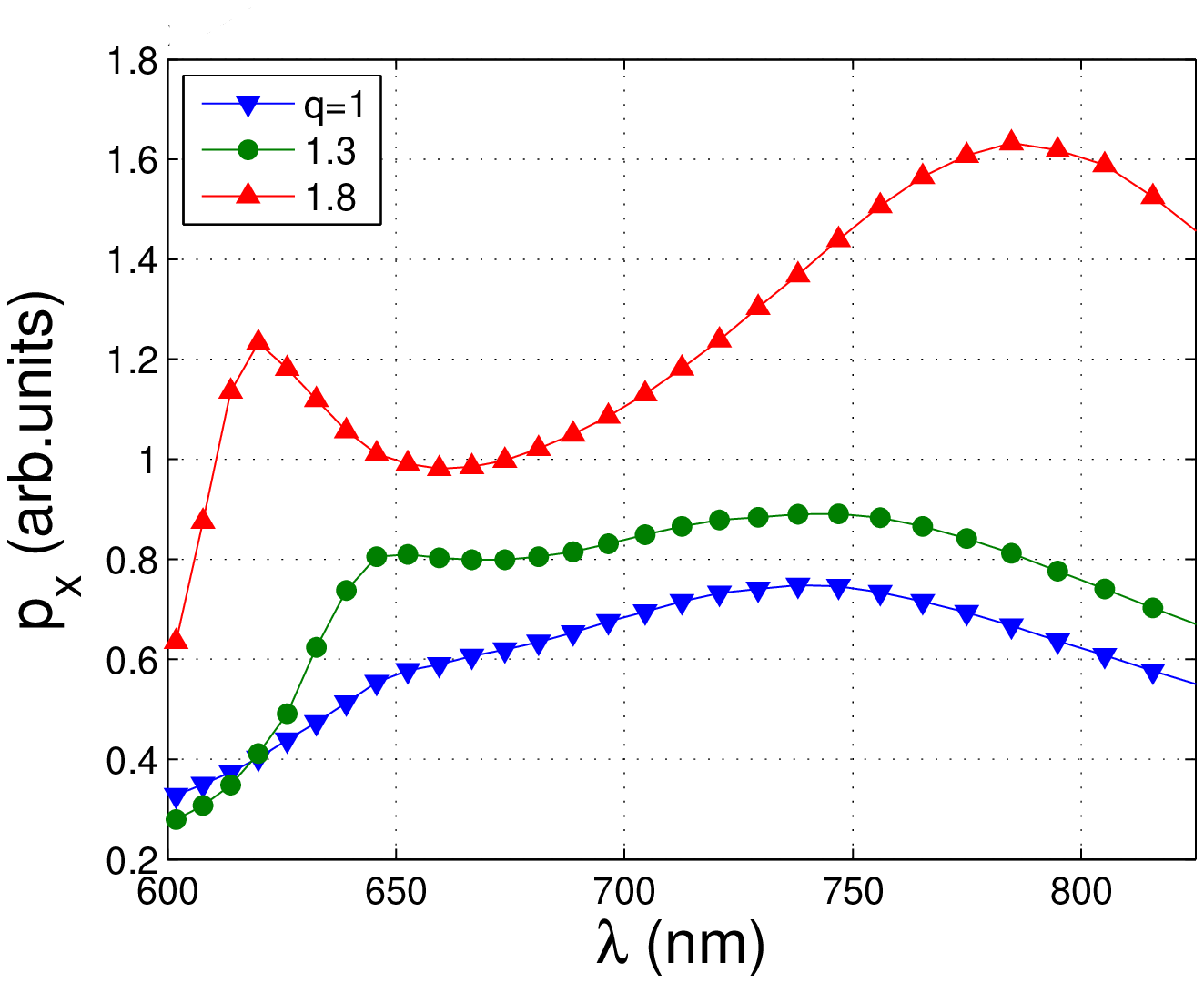}%

(c)\includegraphics[scale=.100,width=17pc,height=13pc]{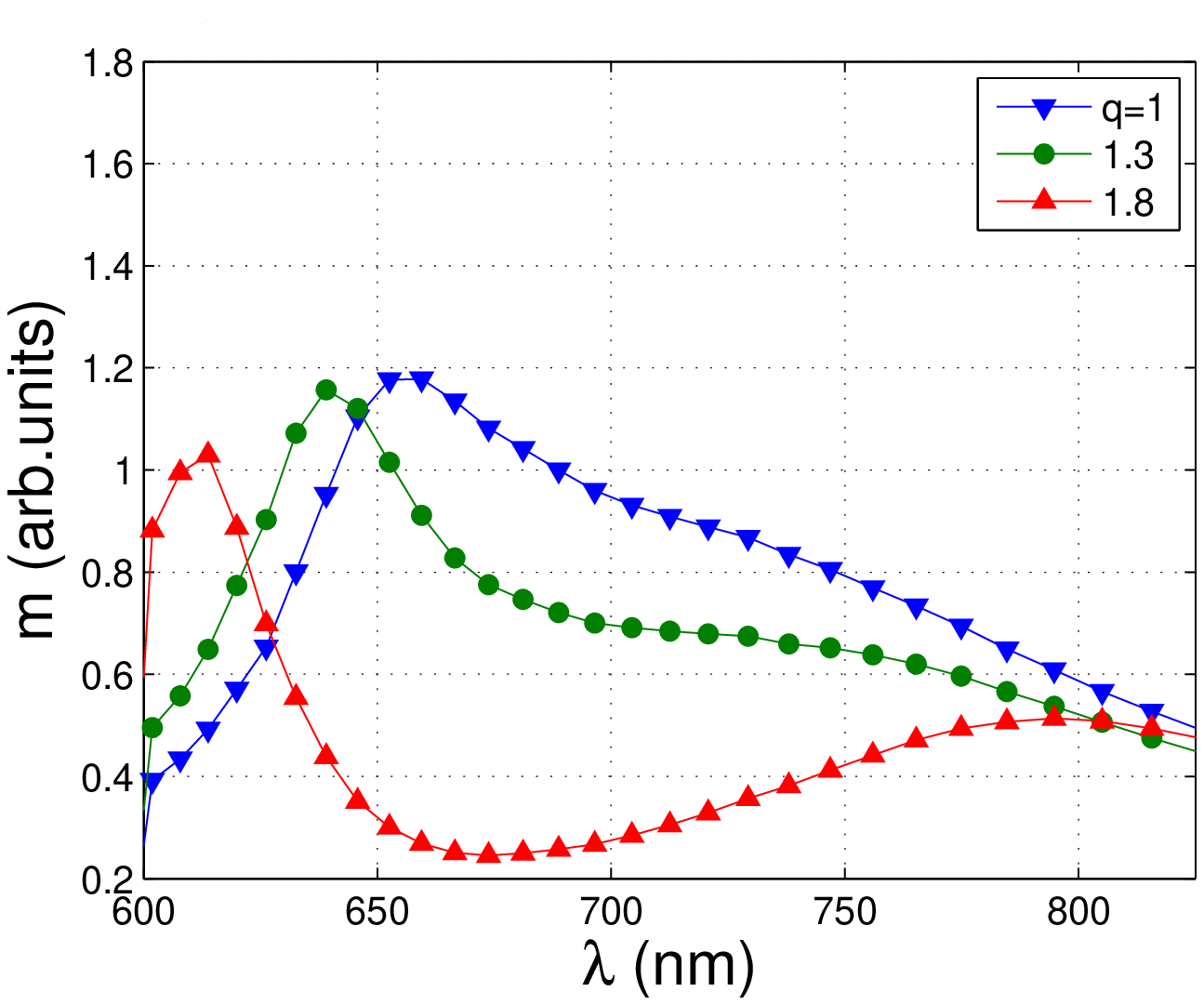}%
\end{center}
\caption{\label{F6}  The magnitudes of the (a) out-of-plane $p_z$
and (b) in-plane $p_x$ components of the ED moment and (c) the
in-plane $m_y$ component of the MD moment of spheroid silicon NPs
irradiated by a SPP plane wave. $q$ is the NP aspect ratio. The
volume of the NPs are corresponded to a spherical NP with the radius
of 95 nm. }
\end{figure}

Weak negative (antiresonant) contribution  of the in-plane ED $p_x$
into the extinction cross section (Fig. \ref{F3}, $\lambda=650$ nm)
is due to the bianisotropic effect, where the ED response is induced mainly
by the electric field generated by the MD and reflected
from the metal substrate surface. This is a reason why this
antiresonance is appeared at the condition of the MD resonance and
the induced in-plane ED moment is out of the phase with external
incident electric field. Generally, the bianisotropic effect can be perceived as
an excitation of the in-plane electric (magnetic) dipole
moment by the magnetic (electric) field of incident electromagnetic
waves.\cite{EvlyukhinACS_Phot2015} In this case the in-plane
components of electric ${\bf p}_{||}=(p_x, p_y)$ and magnetic ${\bf
m}_{||}=(m_x, m_y)$ dipole moments are determined by the expressions
\begin{eqnarray}
{\bf p}_{||}&=&\hat\alpha_e{\bf E}_0+\hat\alpha_{eh}{\bf H}_0\\
{\bf m}_{||}&=&\hat\alpha_h{\bf H}_0+\hat\alpha_{he}{\bf E}_0\:,
\end{eqnarray}
where  the tensors $\hat\alpha_{eh}$, and $\hat\alpha_{he}$
providing the bianisotropy effect and the direct polarizability
tensors $\hat\alpha_e$, $\hat\alpha_h$, are depended on the
electromagnetic interaction between  the NP and substrate. Moreover
$\hat\alpha_{eh}=\mu_0\hat\alpha_{he}$.\cite{EvlyukhinACS_Phot2015}

 In the spectral region corresponding to the $p_x$
antiresonance
 the magnitudes of $p_x$ and $p_z$ are close to each other (Fig.
 \ref{F5}). Therefore both ED components and the MD
  will determine  the scattered light waves  (\ref{EL1}) and
 (\ref{EL2}). However, due to the bianisotropic effect, the scattered electric
 field generated by $p_x$ will be out of phase with respect to the
 electric field generated by $m_y$. As a result the
 scattered  light waves propagating away from the surface
 will be partially suppressed providing the minimum of $\sigma_{\rm Light}$ (Fig. \ref{F2}a)
 and the  maximum  for the ratio
 $\sigma_{\rm SPP}/\sigma_{\rm Light}$ (Fig. \ref{F2}b). Thus the
 spectral difference  between the two resonances presented in Fig.
 \ref{F2}b is explained  $i)$ by  different importance of the in-plane component
 $p_x$ in the two scattering channels and $ii)$ by the bianisotropic effect.

In order to realize the resonant unidirectional and elastic SPP scattering within the same
spectral region, one has  to increase the contribution of the
in-plane component $p_x$ into the elastic SPP scattering. Here, we
suggest to achieve this effect by reshaping spherical particles into
oblate spheroids. It is seen (Fig. \ref{F6}) that different aspect ratios $q$ ($q=1$
corresponds to spherical shape, for oblate spheroids $q>1$) result in
different magnitudes of the dipole components
$p_x$, $p_z$, and $m_y$  excited by SPP plane waves in silicon
oblate spheroids. Note that the considered NPs have the same volume, corresponding to a
sphere with the radius of 95 nm. With increasing of the oblation parameters $q$ the
magnitude of the in-plane dipole component $p_x$ grows including the
two resonances (Fig. \ref{F6}b). The other (out-of-plane) dipole
component $p_z$ is decreased with increasing $q$ (Fig. \ref{F6}a)
providing more important influence of $p_x$ on the SPP scattering in
the case of oblate NPs. Note that the two resonances of $p_x$
($q=1.8$) are appeared at the spectral regions where the
corresponding resonances of $m_y$ are also excited (Fig. \ref{F6}b
and c). This correlation is a result of the bianisotropy effect
connecting the electric and MD resonances. The spectral shifts of
the $p_x$ and $m_y$ resonances with changing of $q$ are a result of
complex combination of the byanisotropy and dressing effects. The
latter can be illustrated using the image theory. For more oblate
spheroid NPs the distance between induced dipoles and their images
(with respect to metal-mirror surface) is decreased providing more
strong interaction between them. As a result the in-plane MD (ED)
resonance could have blue (red) shift because the MD (ED)  and its
image are directed in the same (opposite)
directions.\cite{QuidanACSNano2013}

\begin{figure}
\begin{center}
(a)\includegraphics[scale=.100,width=16pc,height=13pc]{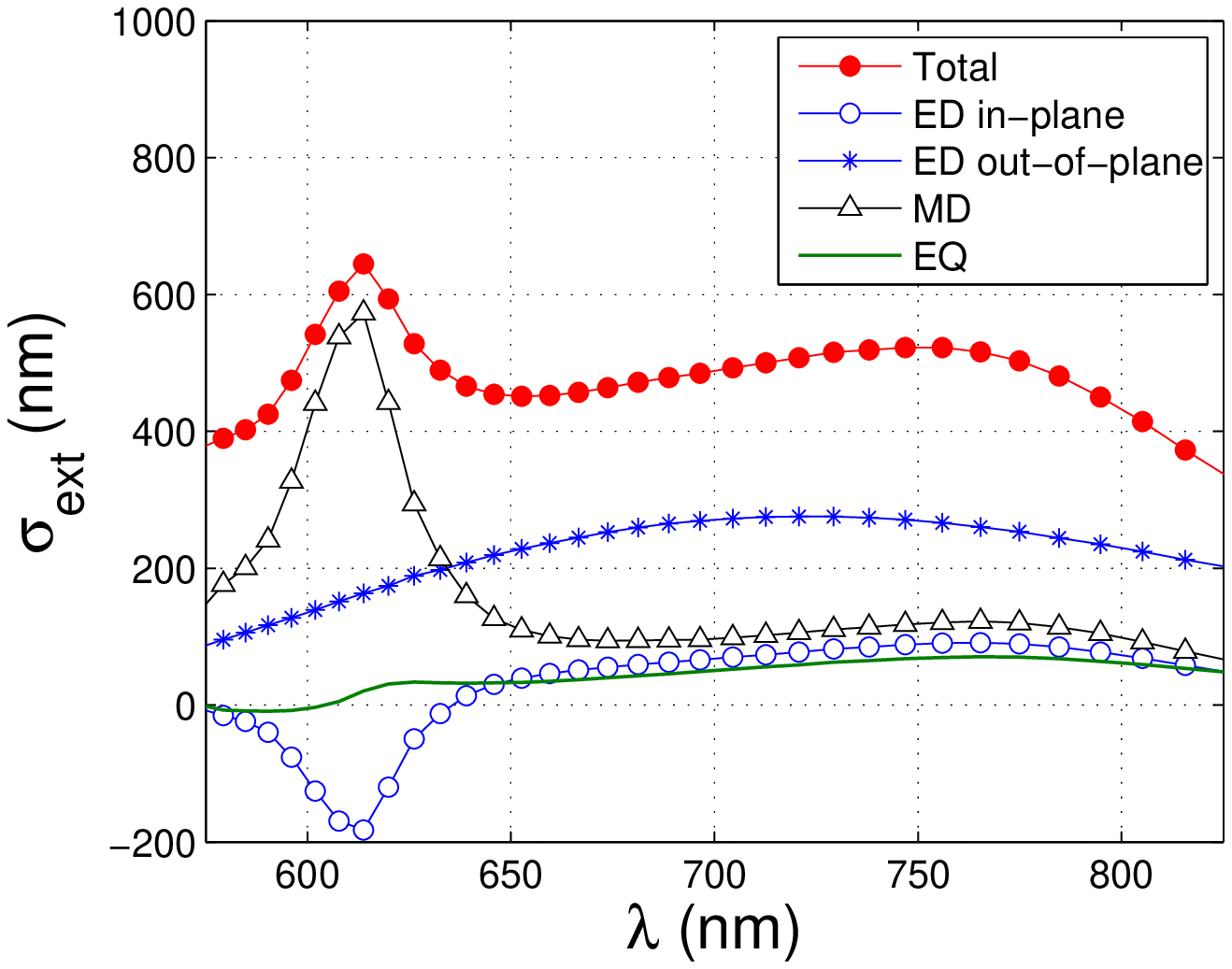}%

(b)\includegraphics[scale=.100,width=16pc,height=13pc]{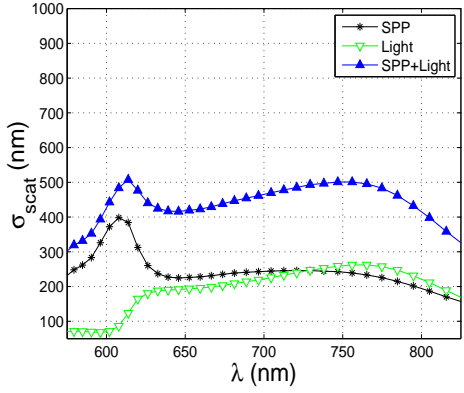}%
\end{center}
\caption{\label{F7} (a) Extinction cross-section spectra of a oblate
spheroid silicon NP (the aspect ratio $q=1.8$, the volume
corresponds to a spherical NP with radius of 95 nm) located on a
gold surface and irradiated by a SPP plane wave. The plot shows
different multipole contributions to the total extinction cross
section: ED electric dipole; MD magnetic dipole; EQ electric
quadrupole. (b) Scattering cross-section spectra of this  spheroid
silicon NP. SPP (Light) corresponds to $\sigma_{\rm SPP}$
($\sigma_{\rm Light}$).}
\end{figure}

\begin{figure}
\begin{center}
\includegraphics[scale=.100,width=17pc,height=13pc]{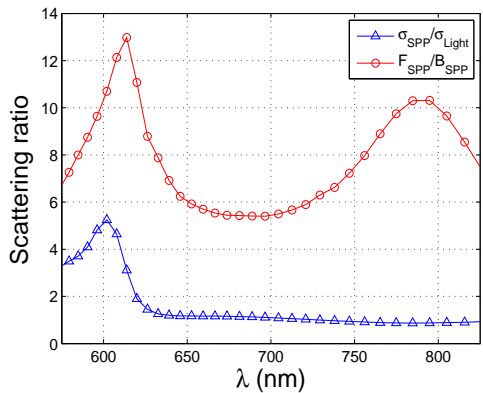}%
\end{center}
\caption{\label{F8} Spectral dependence of the ratios $\sigma_{\rm
SPP}/\sigma_{\rm Light}$ and $F_{\rm SPP}/F_{\rm Light}$ for the
spheroid silicon NP irradiated by a SPP plane wave. The parameters
are as in Fig. \ref{F9}.}
\end{figure}

 Thus the two resonances and the
spectral difference between them shown in Fig. \ref{F6}b and c for
the oblate spheroid NP with $q=1.8$ explain  the spectral features
of the extinction and scattering cross sections presented in Fig.
\ref{F7}. The multipole decomposition in Fig. \ref{F7}a confirms
that the extinction and scattering cross section are determined by
the electric and MD modes excited  by SPP waves inside the oblate
NP. In contrast to the results shown in Fig. \ref{F3} for the
spherical NP the
 negative contribution of $p_x$-term into the short-wavelength resonance of the total extinction
 cross section in Fig. \ref{F7}a is increased indicating on
 importance of the in-plane $p_x$ component in the SPP
 scattering. As a result the elastic SPP
 scattering maximum ($\sigma_{\rm SPP}/\sigma_{\rm Light}$) and the maximum  of the SPP forward
 scattering ($\rm F_{\rm SPP}/\rm B_{\rm SPP}$) are realized at the
 narrow spectral region around $\lambda=610$nm (Fig. \ref{F8}). It
 is seen from the figure that under these conditions the SPP forward (SPP elastic) scattering  is
ten (five) times more effective than the SPP backward (SPP into
light) scattering. The more broad resonance of $\rm F_{\rm SPP}/\rm
B_{\rm SPP}$  at $\lambda=790$nm in Fig. \ref{F8} is realized when
$\sigma_{\rm SPP}/\sigma_{\rm Light}\simeq 1$.

\begin{figure}
\begin{center}
\includegraphics[scale=.100,width=17pc,height=12pc]{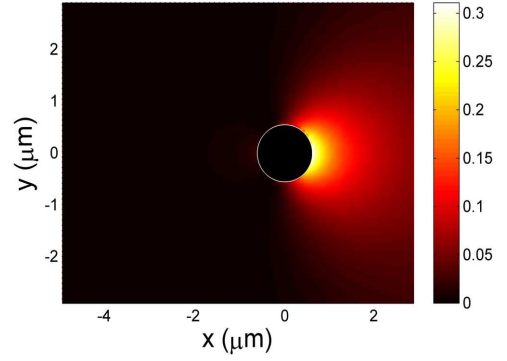}%

\end{center}
\caption{\label{F9} Scattered SPP electric field intensity
(arb.units) calculated 150 nm above the air-gold interface for the
SPP plane wave (light wavelength 614 nm) being incident on the
oblate spheroid silicon NP ($q=1.8$).  The black circles indicate
the NP position. }
\end{figure}

\begin{figure}
\begin{center}
\includegraphics[scale=.100,width=18pc,height=11pc]{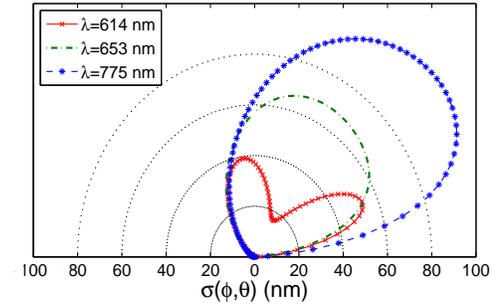}%
\end{center}
\caption{\label{F91} Differential scattering cross-sections
(scattering diagrams)  of SPP-into-light scattering  by  the oblate
spheroid silicon NP ($q=1.8$) calculated in the $xz$-plane. }
\end{figure}

For the explicit demonstration of the SPP backward scattering
suppression, we calculated  the SPP intensity distribution in a plane
located in near-field zone above the metal substrate surface with the
oblate NP ($q=1.8$) illuminated by an incident SPP wave corresponding
to the resonant wavelength $\lambda=614$nm. The spatial intensity distribution of
scattered SPP demonstrates close to complete suppression
of the SPP backward scattering (Fig. \ref{F9}). Moreover, the SPP-into-light
scattering is also significantly weakened at this resonant
wavelength (Fig. \ref{F91}). Concluding this
subsection, we note that, if a space gap between the oblate NPs and
metal substrate surface increases, the spectral distribution of the
SPP scattering cross section begins to look like that of the scattering
cross section for spherical NPs located on the substrate surface.

\begin{figure}
\begin{center}
(a)\includegraphics[scale=.100,width=17pc,height=13pc]{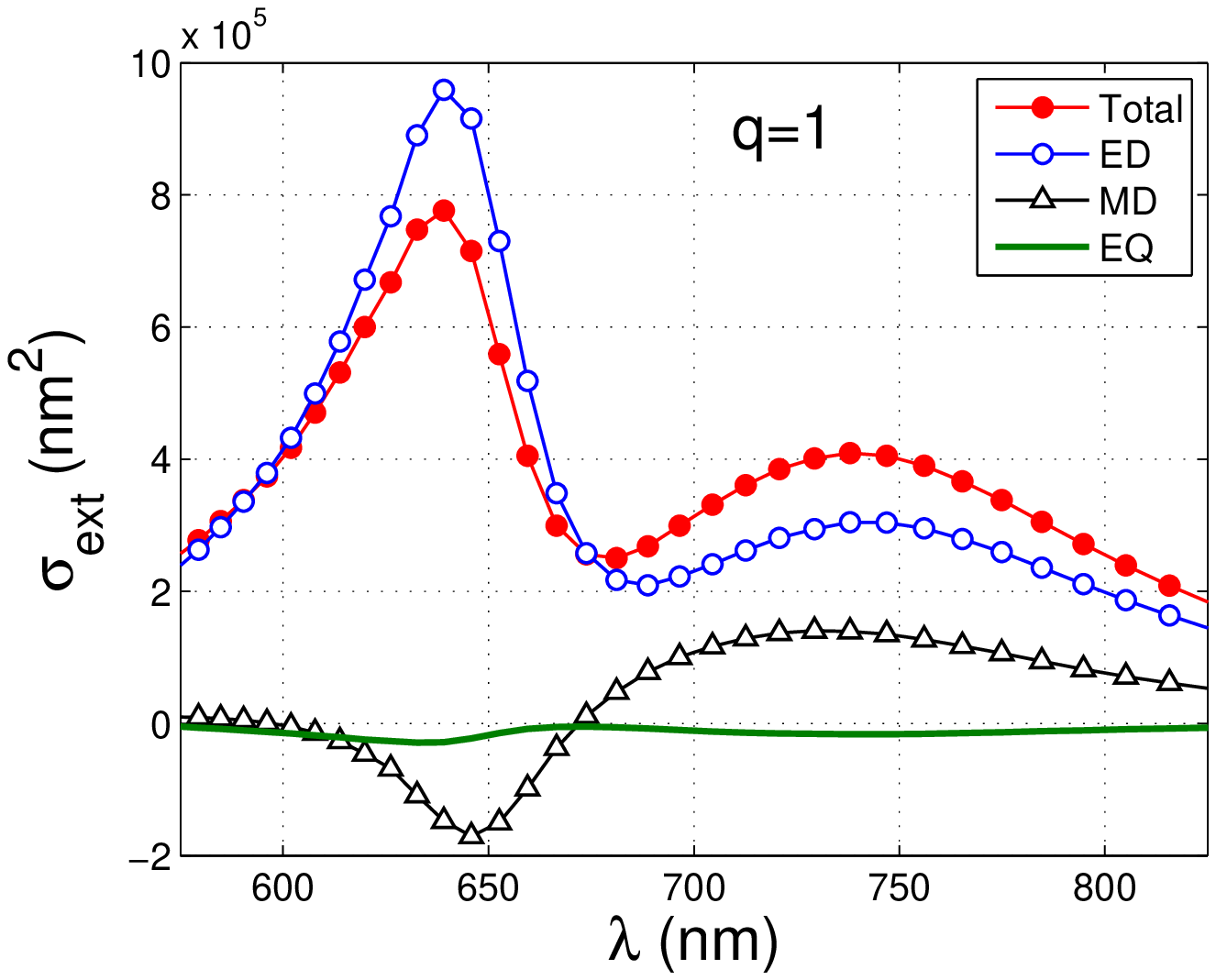}%

(b)\includegraphics[scale=.100,width=17pc,height=13pc]{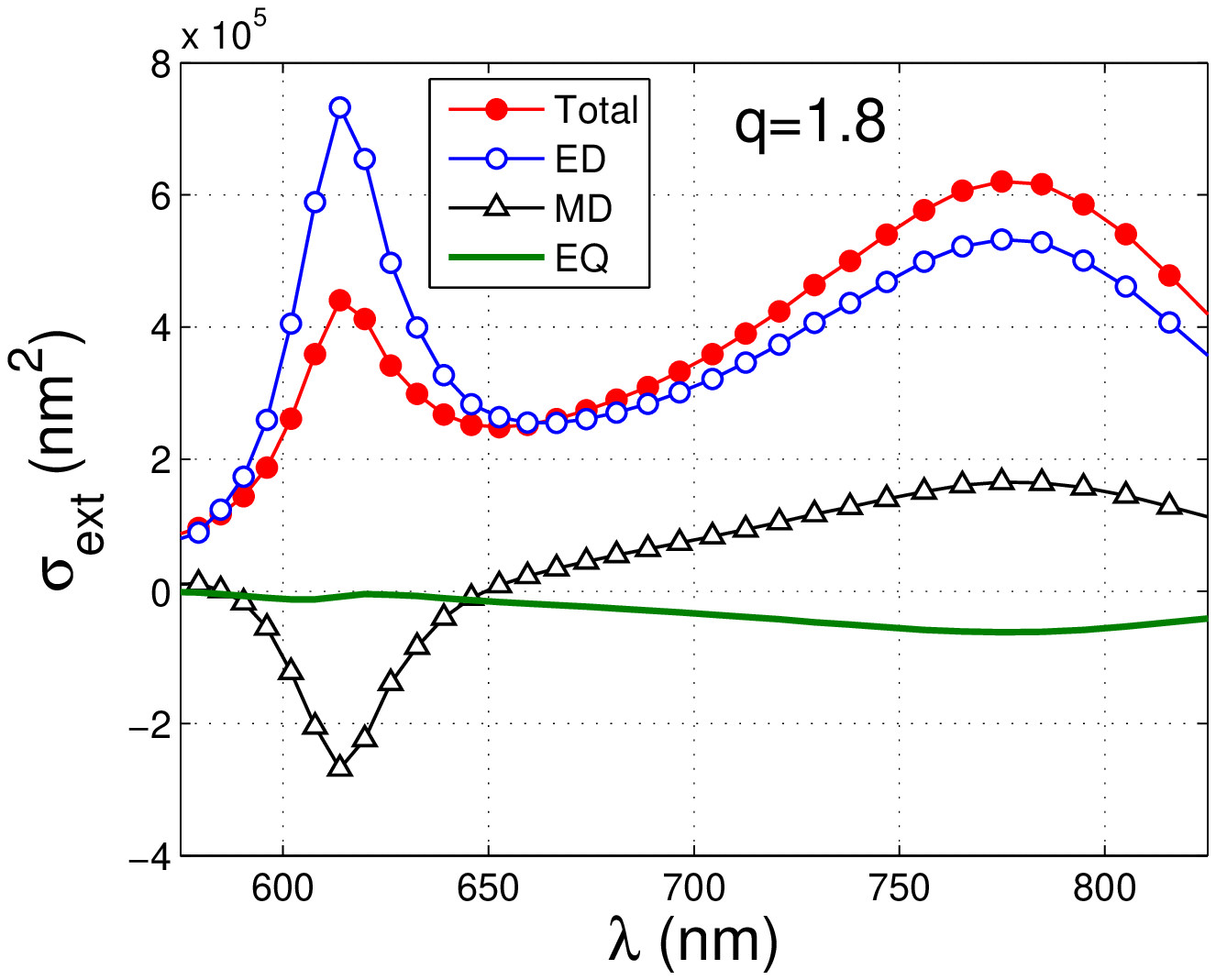}%

(c)\includegraphics[scale=.100,width=17pc,height=13pc]{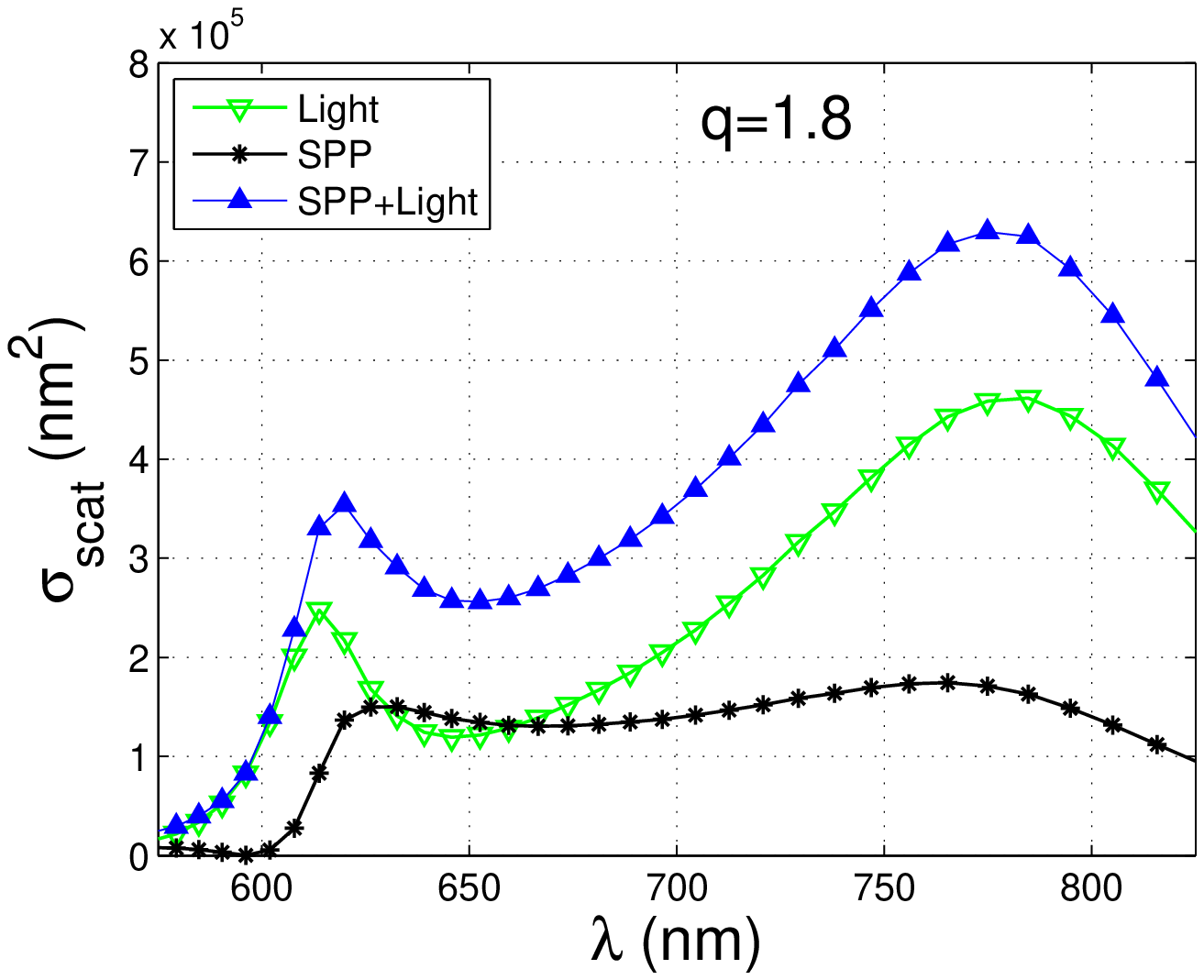}%
\end{center}
\caption{\label{F11}  Extinction cross-section spectra of (a)
spherical and (b) oblate spheroid silicon NPs ($q$ is the aspect
ratio, the volumes correspond to a spherical NP with radius of 95
nm) located on a gold surface and irradiated by a normally incident
light plane waves. The plot shows different multipole contributions
to the total extinction cross section. (c) Scattering cross-section
spectra of the same spheroid silicon NP. SPP (Light) corresponds to
$\sigma_{\rm SPP}$ ($\sigma_{\rm Light}$). }
\end{figure}

\begin{figure}
\begin{center}
\includegraphics[scale=.100,width=16pc,height=13pc]{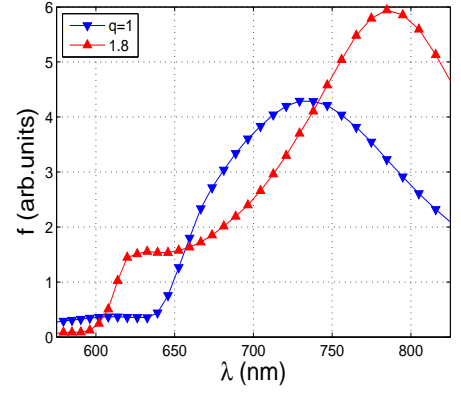}%
\end{center}
\caption{\label{F12} Spectral dependence of the function
$f=|(iap_x-(1-a^2)\sqrt{\mu_0\varepsilon_0}\:m_y)|^2$ for
 the spherical and oblate spheroid silicon  NPs,  Eq. (\ref{ES}). }
\end{figure}

\begin{figure}
\begin{center}
\includegraphics[scale=.100,width=16pc,height=13pc]{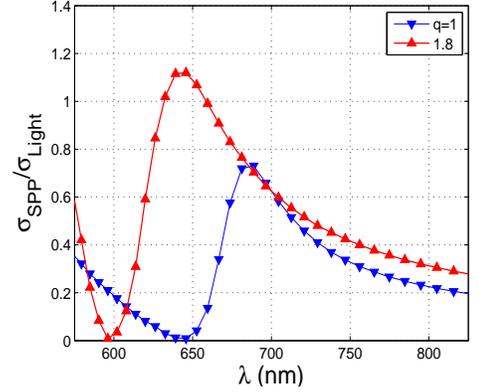}%
\end{center}
\caption{\label{F13} Spectral dependence of $\sigma_{\rm
SPP}/\sigma_{\rm Light}$ calculated for  the spherical and oblate
spheroid silicon  NPs irradiated by normal-incident light plane
waves. }
\end{figure}

\begin{figure}
\begin{center}
(a)\includegraphics[scale=.100,width=17pc,height=13pc]{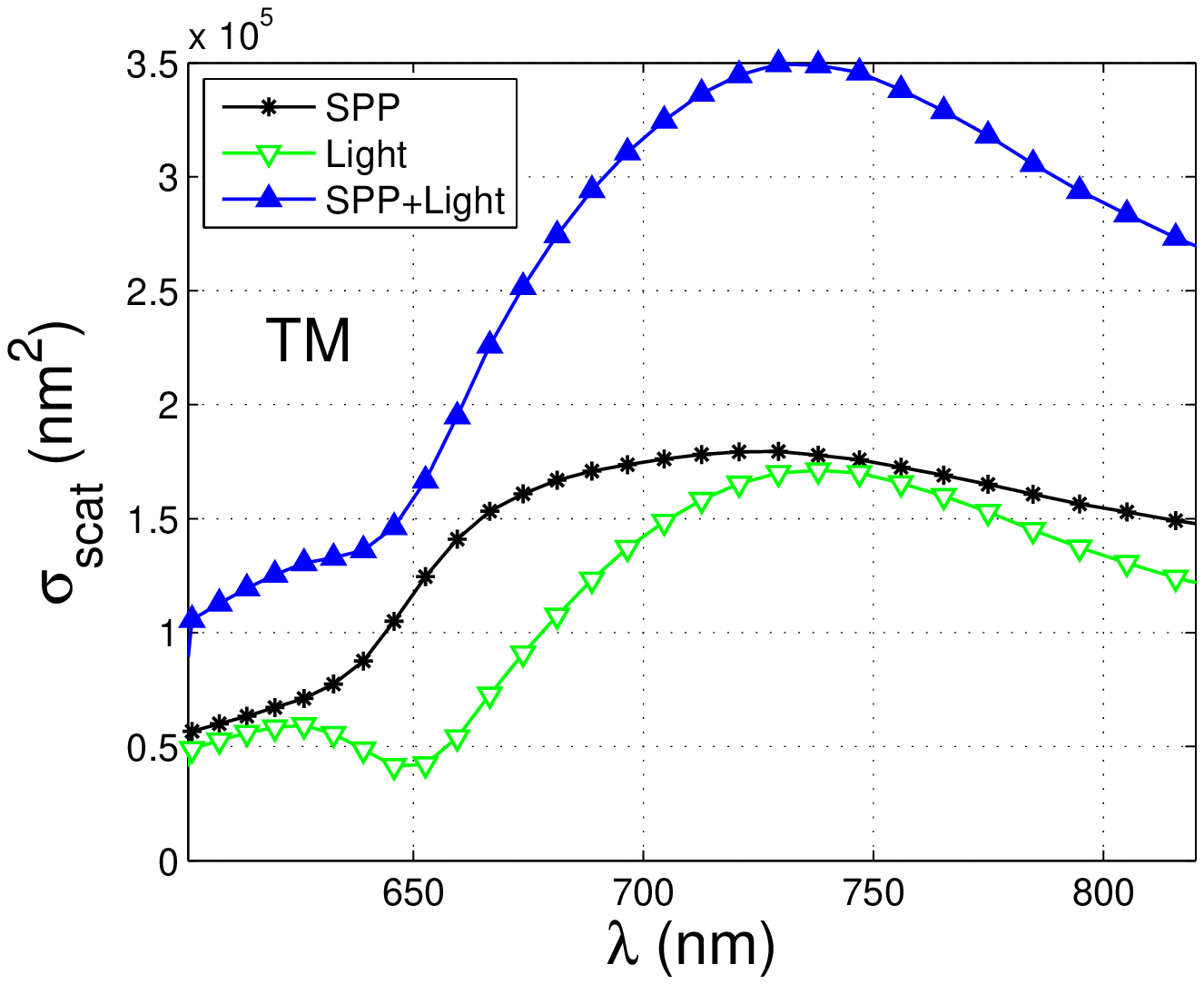}%

(b)\includegraphics[scale=.100,width=17pc,height=13pc]{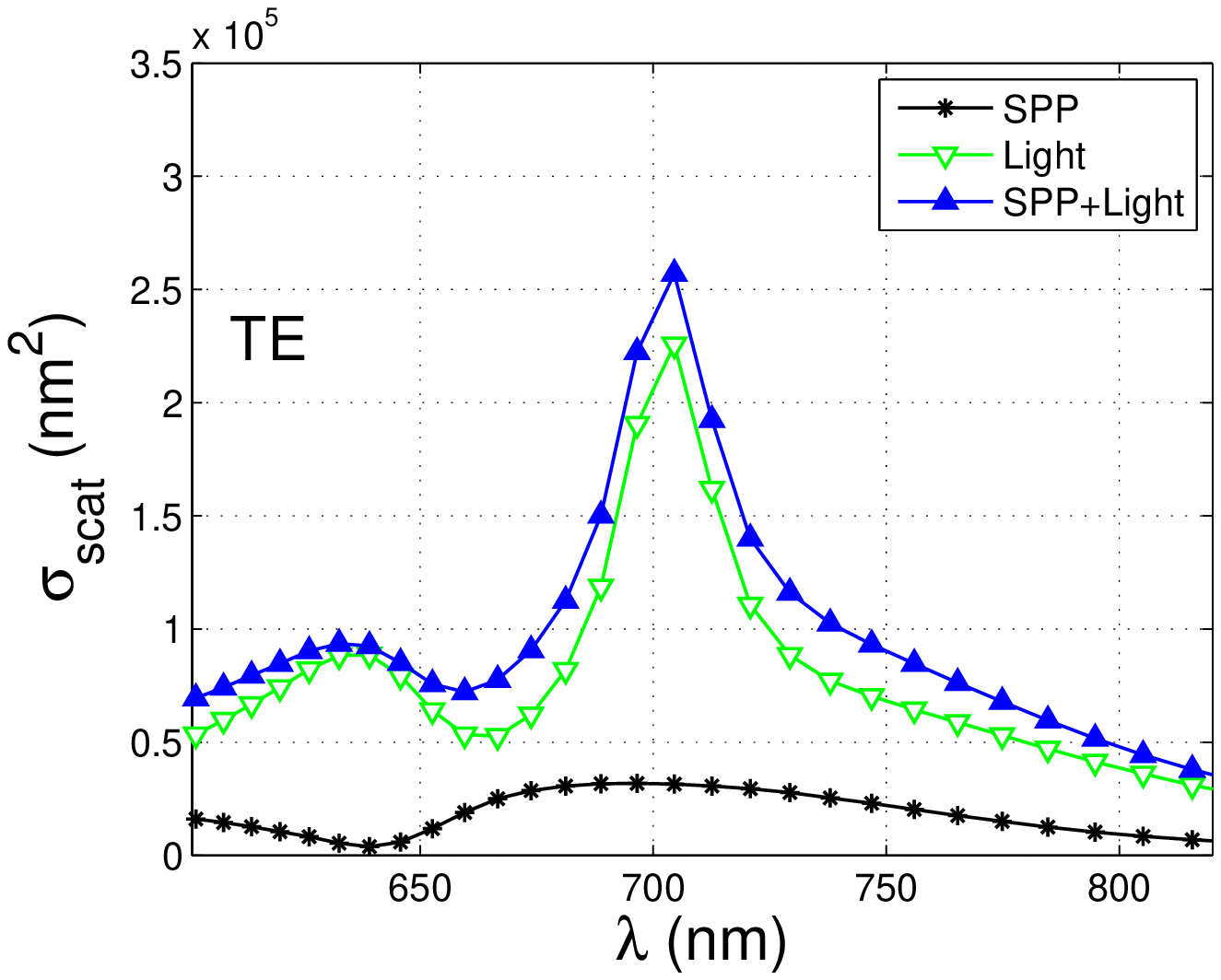}%
\end{center}
\caption{\label{F14} Scattering cross-section spectra of  spherical
 silicon NPs with the radius of 95 nm located on a gold surface and irradiated by a
incline-incident light plane waves. The incident angle is equal to
$70^{\rm o}$. SPP (Light) corresponds to $\sigma_{\rm SPP}$
($\sigma_{\rm Light}$). }
\end{figure}

\subsection{Light scattering}

Let us consider the reciprocal light scattering from silicon NPs
located on a gold surface and having parameters as in the previous
subsection. Here we assume a normal incidence of linear-polarized
light plane waves as it is shown in Fig. \ref{F1}b. The spectra of
extinction cross sections for spherical and oblate spheroid NPs are
presented  in  Fig. \ref{F11}. From the multipole decompositions
included there it is clear that the optical properties of the NPs
are determined by solely their electric and MD moments. Due to the
byanisotropy effect the both resonant peaks of the extinction cross
sections in Fig. \ref{F11}a,b are corresponded to the simultaneous
resonant excitation of the electric and MD moments. In contrast to
the case of the SPP scattering here the partial negative
contribution in total extinction cross sections is corresponded  to
the MD term (MD) of the multipole decomposition (Fig. \ref{F11}).
That is resulted from different space distributions  of the electric
and magnetic fields of the incident SPP and light waves near a metal
surface. Note that the NP MD moment is not in phase with the
magnetic field of the incident light waves at the spectral region
where the MD contribution in the extinction cross section is
negative. Another important difference from the case of SPP
scattering is the absence of the out-of-plane component  $p_z$ under the
condition of normal light incidence. Consequently, Eq.(\ref{ESPPZ}) reduces to the following one:
\begin{equation}\label{ES}
E_{{\rm SPP}z}\sim
(iap_x-(1-a^2)\sqrt{\mu_0\varepsilon_0}\:m_y)\cos\varphi\:
\end{equation}
for the incident light polarized along the $x$-axis. As
a result, we obtain the condition for the suppression of SPP
excitation in the system
\begin{equation}\label{con}
p_x\approx-\sqrt{\frac{\mu_0\varepsilon_0\varepsilon_s}{\varepsilon_d}}\:m_y\:,
\end{equation}
where $\varepsilon_s$ is the complex number and  one takes
$\sqrt{-1}=i$. Owing to the bianisotropy interaction  existing only
between the in-plane ED and MD components, the
condition (\ref{con}) could be satisfied in the spectral region,
where resonances of the ED and MD contributions in the extinction
spectrum have opposite signs. Indeed, one notices (Fig. \ref{F11}c) that the
scattering of light into SPP is significantly suppressed in the
spectral region below $\lambda=600$ nm. Figure \ref{F12}
demonstrates the spectral dependence of the expression
$f=|(iap_x-(1-a^2)\sqrt{\mu_0\varepsilon_0}\:m_y)|^2$ from
(\ref{ES}) for the spherical and spheroid NPs. In the both cases the
minimum (maximum) of excited SPPs  in the considered systems is
realized at the condition of the short-wavelength (long-wavelength)
resonance of the total extinction and scattering cross sections
(Fig. \ref{F11}). Importantly the SPP suppression is realized for
silicon nanospheroids with different aspect ratios at the condition
of the short-wavelength resonance because  the NP electric and MD
resonances  are automatically adjusted to each other due to the
strong bianisotropy effect. Changing the NP shape  it is
 possible to shift the spectral point where the SPP excitation is
 suppressed (Fig. \ref{F13}). Similar behavior can be also
 realized by changing the NP sizes at the fixed shape.

Concluding this section let us briefly discuss the scattering of
incline-incident light plane waves. Figure \ref{F14} demonstrates
the scattering cross sections calculated for the $70^{\rm o}$
incidence of light waves with TM- and TE-polarization. In the case
of the TM-polarization light (Fig. \ref{F14} a) the partial
scattering cross section $\sigma_{\rm SPP}$ exceeds the section
$\sigma_{\rm Light}$ because of the contribution of the large
out-of-plane ED component $p_z$ excited by the electric field of the
incident light wave. Partially the situation reminds the SPP
scattering (Fig. \ref{F2} a) where the out-of-plane ED component
$p_z$ plays a principal role.  In the case of the TE-polarization
light (Fig. \ref{F14} b) the efficiency of scattering into light
exceeds the scattering into SPPs as it is realized in the case of
the normal-incident light. Additional resonant peak at $\lambda=710$
nm in Fig. \ref{F14} b corresponds to resonant excitation of  the
out-of-plane MD component $m_z$. However, as already has been
indicated above, this component does not participate in the SPP
excitation and can increase the only light-into-light scattering
cross section.

\section{Conclusion}

The multipole analysis of SPP and light scattering by
arbitrary-shaped NPs located on a plane surface has been applied to
study the optical response of individual spheroidal silicon NPs with
sizes of the order of 200 nm being placed near a gold surface. Using
numerical calculations by the DDA method  together with the
multipole decomposition procedure, we showed that the extinction and
scattering cross sections of the silicon NPs have two resonances (in
the spectral range of $\lambda \in [575-825]$ nm) corresponding to
the resonant excitations of their ED and MD moments. Due to the
bianisotropy effect, the in-plane ED and MD components are excited
by both electric and magnetic fields of incident electromagnetic
waves. As a result, the resonant peaks of the extinction  cross
sections can include both ED and MD resonant contributions. It was
found that, in the case of SPP scattering by oblate spheroidal
silicon NPs, the ED and MD resonant excitations can ensure resonant
unidirectional and elastic SPP scattering within the spectral range
that is determined by the NP aspect ratio. For NPs with the aspect
ratio of 1.8, it was obtained that at the wavelength $\lambda=610$
nm the SPP forward (elastic)scattering is ten (five) times more
effective than the SPP backward (SPP into light) scattering. The
suppression of SPP scattering into light is connected with the
destructive interference between light waves radiated by the induced
ED and MD moments of NP.

We have also discussed a role of the bianisotropy effect in the
realization of the efficient elastic SPP scattering. It was shown
that due to the bianisotropy  effect the
\begin{figure}
\begin{center}
(a)\includegraphics[scale=.100,width=17pc,height=13pc]{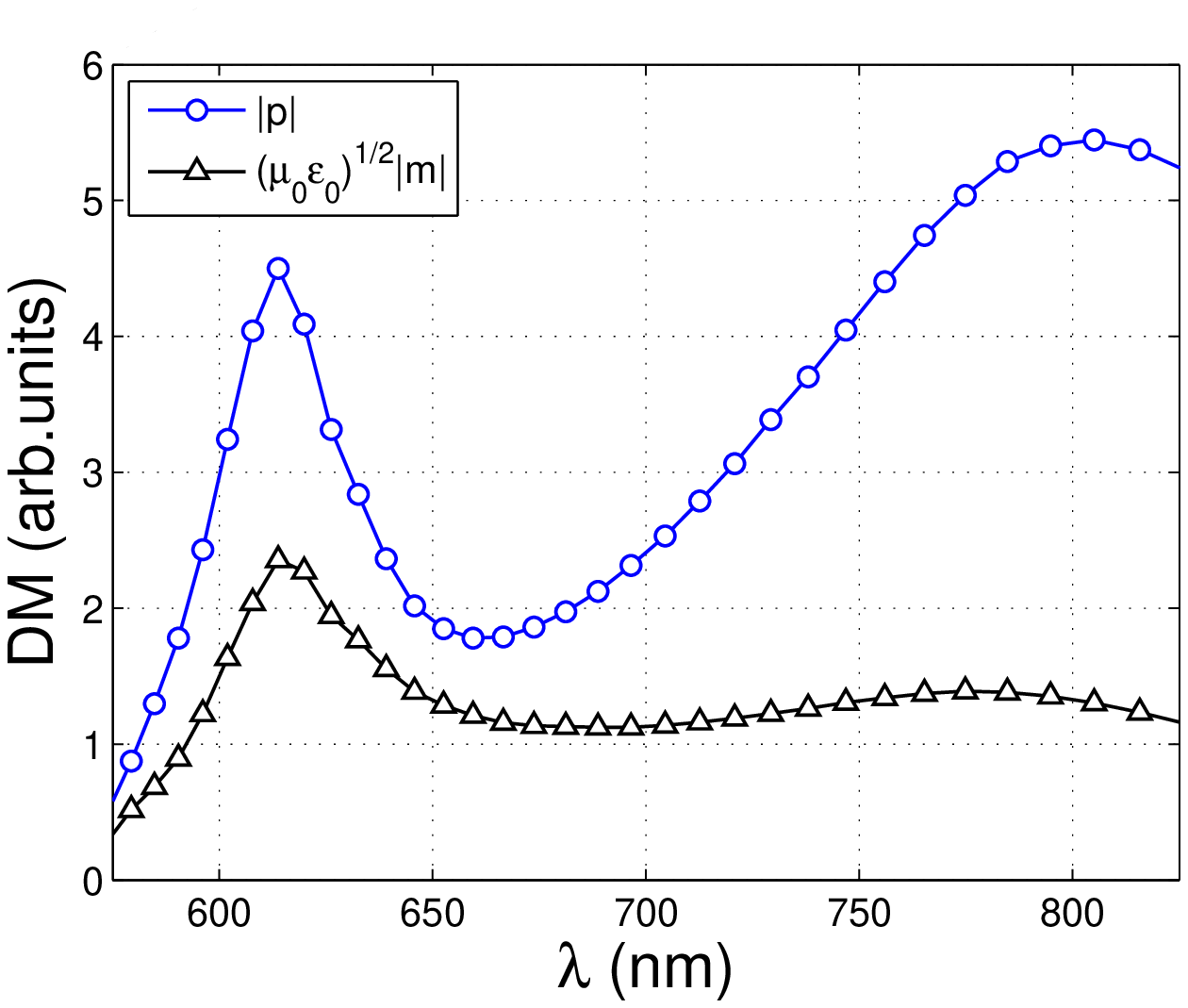}%

(b)\includegraphics[scale=.100,width=17pc,height=13pc]{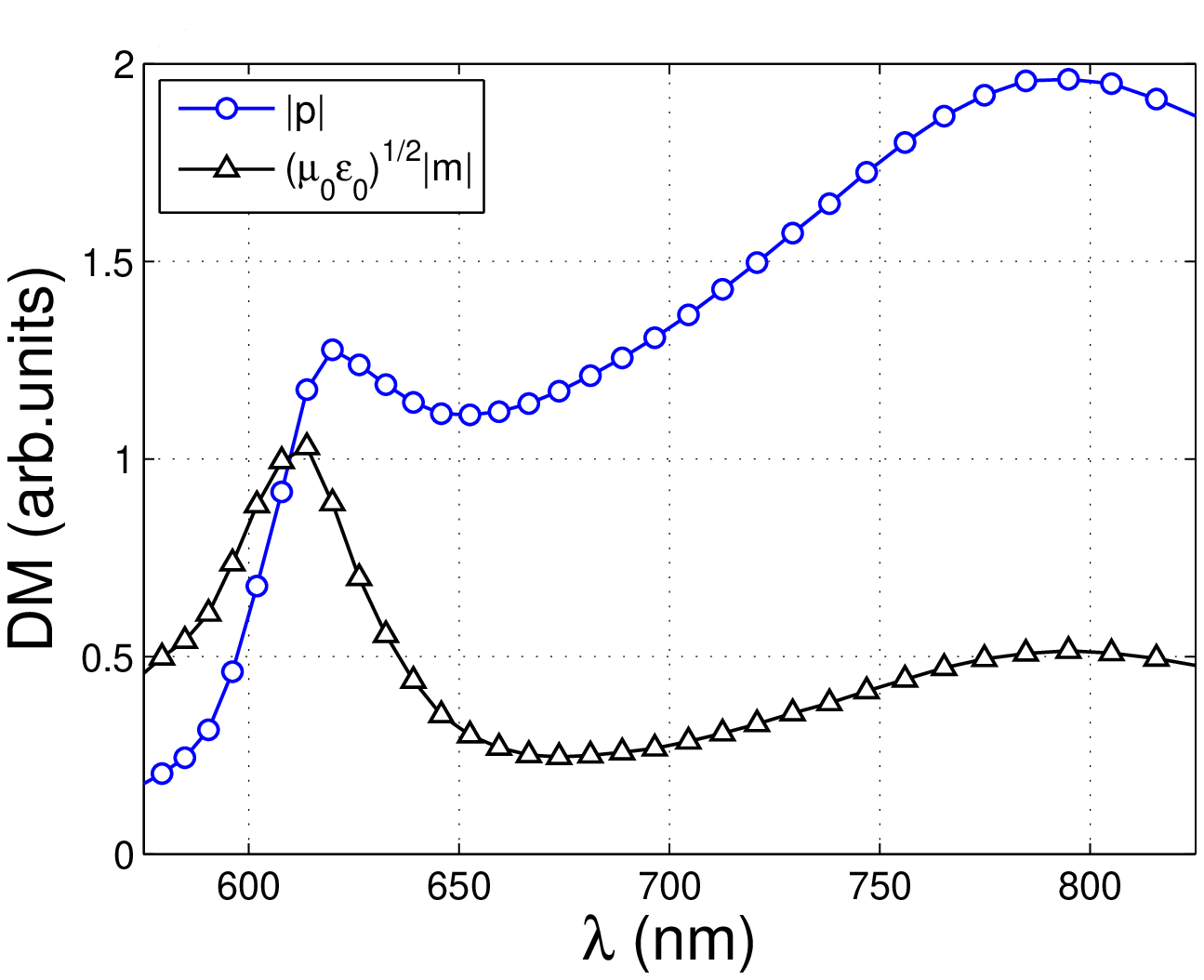}%
\end{center}
\caption{\label{F15} Magnitudes of the ED moment $\bf p$ and the MD
moment $\sqrt{\mu_0\varepsilon_0}\:\bf m$ of a spheroidal silicon NP
with $q=1.8$ irradiated by (a) normal-incident light plane waves and
(b) by  SPP plane waves.  The NP volume corresponds to that of a
spherical 95-nm-radius NP.}
\end{figure}
contributions of the in-plane ED and MD components to the extinction
cross section can be of opposite signs. In this case, the
destructive interference between the waves generated by these dipole
components can suppress the considered scattering channel. In the
case of SPP incidence, this destructive interference practically
eliminates (out-of-plane) scattering into light, whereas in the case
of normally incident light, this ensures efficient suppression of
SPP excitation in the system. The appropriate analytical condition
for the suppression of the SPP excitation has been obtained.
Difference in SPP and light scattering by NPs is connected with
different spatial distributions of the electric and magnetic fields
created in these two cases near a gold surface. In the case of
normally incident light waves, the induced ED $\bf p$ exceeds the
induced
 MD $\sqrt{\mu_0\varepsilon_0}\:\bf m$ (where
 $\sqrt{\mu_0\varepsilon_0}$ is the dimension factor) within the considered spectral range (Fig.
 \ref{F15} a). As a result, only the SPP scattering suppression can be realized
 in the system at the short-wavelength resonance (Fig. \ref{F15} a).
 In the case of SPP scattering, the corresponding magnitudes
 of $\bf p$ and $\sqrt{\mu_0\varepsilon_0}\:\bf m$ can be equal to each other at
 the short-wavelength resonance
 (Fig. \ref{F15} b), providing the possibility of suppression of SPP-into-light
 scattering  (Fig. \ref{F8}).
 Additionally, it was shown that, in the case of oblique light incidence, the
 scattering cross sections for scattering into
 light and SPP excitation are strongly dependent on the light polarization.

Summarizing, the results obtained demonstrate directly that
high-refractive-index dielectric NPs (for example, crystalline
silicon NPs), supporting resonant excitation of the MD and ED modes
and being located on metal (gold, silver)) substrates, can be used
for the realization of unidirectional and elastic SPP scattering as
well as for the suppression of SPP scattering when illuminated by
light. We believe that these unique optical properties are very
important for designing SPP-based photonic components and
metasurfaces for control and manipulation of SPP and light waves.

{\bf ACKNOWLEDGEMENT}

\noindent We acknowledge financial support for this work from the
University of Southern Denmark (SDU 2020 funding) and from the
European Research Council, Grant No. 341054 (PLAQNAP).

\end{document}